\documentclass[onecolumn,notitlepage,amsmath,amssymb,aps,pre,final]{revtex4-2}

\renewcommand{\exp}[1]{\mathrm{e}^{#1}}
\usepackage{mathtools}
\usepackage{graphicx,xcolor}
\usepackage{dcolumn}
\usepackage{bm}
\usepackage{hyperref,cleveref}
\crefname{section}{Sec.}{Secs.}
\crefname{subsection}{Sec.}{Secs.}
\crefname{subsubsection}{Sec.}{Secs.}
\crefname{appendix}{App.}{Apps.}
\crefname{subappendix}{App.}{Apps.}
\hypersetup{colorlinks=true,linkcolor=red!80!blue,citecolor=green!60!blue,filecolor=cyan!80!blue,urlcolor=magenta!80!blue}
\usepackage{nicefrac}
\newcommand{\bigcomma}{\mathpunct{\scalebox{1.5}{,}}}

\usepackage{lmodern}
\let\vec\bm
\DeclareMathOperator{\grad}{\vec{\nabla}\!}

\newcommand{\op}[1]{\mathsf{#1}}
\newcommand{\Jc}{\operatorname{J}_\text{coll}}
\newcommand{\Jm}{\operatorname{J}^\text{MF}_\text{coll}}
\newcommand{\Jl}{\operatorname{J}^\text{L}_\text{coll}}
\newcommand{\Jd}{\operatorname{J}^\text{D}_\text{coll}}
\newcommand{\Jda}{j_1}
\newcommand{\Jdb}{j_2} 
\newcommand{\Jdc}{j_3}
\newcommand{\tildeJdc}{\tilde{\jmath}_3}
\newcommand{\Pe}{\mathrm{Pe}}
\newcommand{\opd}{\!\operatorname{d}\!}
\newcommand{\ii}{\mathrm{i}}

\let\eps\varepsilon
\let\epsilon\varepsilon
\let\theta\vartheta
\newcommand{\abs}[1]{\lvert #1\rvert}
\DeclareMathOperator{\sgn}{sgn}
\DeclareMathOperator{\Res}{Res}

\begin{document}

\title{Statistical mechanics of noisy collisions in active matter: Dyson-Landau theory}

\author{Horst-Holger Boltz}
\email[]{horst-holger.boltz@uni-greifswald.de}

\author{Max-Robin Schultz}
\author{Thomas Ihle}

\affiliation{Institute for Physics, Greifswald University,
17489 Greifswald, Germany}

\date{\today}

\begin{abstract}
Active systems and other coarse-grained descriptions are often modeled by generalized dynamics, effective equations of motion that are not bound by the constraints of genuinely microscopic physics. In particular, these equations are generally stochastic on the smallest resolved scale. This poses a challenge to first-principles statistical physics, which usually derives emergent noise on larger scales from deterministic interactions. Here, we show how to systematically incorporate microscopic noise into kinetic theory in the limit of weak noise, i.e., high activity. By means of a Dyson expansion of the noisy free propagator, we derive the leading correction to the Landau kinetic equation for weakly interacting systems. We find that the common practice of combining a deterministic collision integral with the diffusive terms of the Fokker--Planck equation is the correct leading order. For dry aligning active matter, however, the corrections are non-analytic in the noise strength, as nearly parallel particles interact for arbitrarily long times. For noise that acts only during interactions, a minimal model of stochastic interactions, noise and collisions do not decouple: the leading corrections are odd in the coupling strength, partly logarithmically enhanced, and amount to an orientation-dependent effective noise strength. Agent-based simulations confirm the predicted dependence on the sign of the coupling. Even in this minimal model, the effect of noise on collisions can thus not be inferred from symmetry alone.
\end{abstract}

\maketitle

\section{Introduction}

A staple of modern science in many contexts is agent-based modeling~\cite{schweitzer2003,bonabeau,castellano}. A typical scale separation of time and length scales allows for a division of a system into (often identical) agents that have some dynamics in space and interactions with other agents. Internal degrees of freedom, accessible or not, are not resolved explicitly. Thus, the character of the equations of motion is that of effective, coarse-grained models. As such, they are not subject to the same constraints that are valid for genuinely microscopic dynamics. These generalized dynamics are in general non-Hamiltonian, they generally conserve neither energy nor momentum, irrespective of the symmetries of the model and, particularly relevant for this work, they can include stochastic contributions.

This broader variety of dynamics poses a non-trivial challenge for statistical physics. Without an underlying Hamiltonian there is in general no conserved energy, no Gibbs-Boltzmann distribution to postulate, no entropy or free energy to be extremized, and no detailed balance. The noise is not tied to a temperature, so there is also no direct fluctuation-dissipation relation to lean on~\cite{loos2020,fodor2022}. The usual starting points of equilibrium statistical mechanics are therefore unavailable, and the statistical description has to be built from the equations of motion directly. One approach to statistical mechanics that has no prerequisites in model structure is kinetic theory~\cite{balescu,kreuzer,cercignani1993,villani2002,cercignani2012,pareschi,dorfman2021}, the systematic derivation of an ensemble-level description from the actual, model-specific equations of motion.

A widespread approach in this vein is mean-field theory, which is based on the assumption of molecular chaos, i.e. the statistical independence of the agents. This is asymptotically correct for weak interactions and in the fully connected (every agent interacting with every other) limit. On the mean-field level, there are no discrete interaction events and the influence of intrinsic noise on the statistical theory is rather trivial: it gives rise to diffusive terms in the Fokker--Planck equation which are conceptually unaffected by the mean-field closure and are, thus, mutatis mutandis present in the resulting Vlasov-like equation for the one-particle (probability density) function. Mean-field theory can fail both quantitatively and qualitatively. For example, there are cases of cohesion-type and vision-based models of self-propelled active particles that have an additional rotational symmetry leading to vanishing mean-field forces \cite{negi2022,loos2023,negi2024,das2024,shea2025,huang2025}.

The systematic next step is to account for correlations built up during interactions under the assumption of one-sided molecular chaos: particles are treated as independent at the onset of a collision. Tracking the build-up of correlations during the collision event requires evaluation of two-particle dynamics. If this is done by means of the full two-particle propagator, which amounts to a binary collision approximation, one can derive a Boltzmann equation~\cite{kreuzer,ihle2023}. Its collision integral can rarely be evaluated analytically. More generally accessible is the kinetic approach leading to a Landau equation~\cite{landau1936,balescu}, which is perturbative in the interaction and uses the free two-particle propagator.

In all of these treatments the intrinsic noise enters only as a separate diffusion term acting on the one-particle function, decoupled from the collision operator, and the collision integral itself is built from noiseless, deterministic trajectories \cite{patelli_2021,boltz2024,soto2024,pinto2025,ihle2023,pinto2026}. For weak noise, where interactions remain well-defined collision events, we are not aware of any work that treats the propagation of microscopic stochasticity through the collision integral itself, that is, the interplay of noise and interaction in continuous-time dynamics. Progress in this direction has been made for discrete dynamics in which interactions are instantaneous~\cite{kapral1999,*ihle2001,ihle2011}, and for hard-core active Brownian particles, for which continuum theories have been derived by matched asymptotics~\cite{bruna2022,*kalz2024} in the opposite limit, where diffusion dominates the encounter. This reflects a more general imbalance: the limit of low activity, i.e., of strong noise, is accessible by expansions around equilibrium and has received considerable attention~\cite{freidlinwentzell,fodor2016,vrugt2023}, whereas the limit of weak noise, in which interactions are distinct collision events, has been studied much less~\cite{boltz2025}. The historical direction of kinetic theory is the inverse: to derive the effective noise of coarse-grained theories from deterministic microscopic dynamics. A prominent example is Hilbert's sixth problem concerning the axioms of physics~\cite{hilbert}, which has led to a program of deriving the hydrodynamic equations from hard-sphere collisions via the Boltzmann equation~\cite{lanford1975,gallagher2013,*bodineau2016,deng2025a,*deng2025b}.

\begin{figure}[bth]
\includegraphics{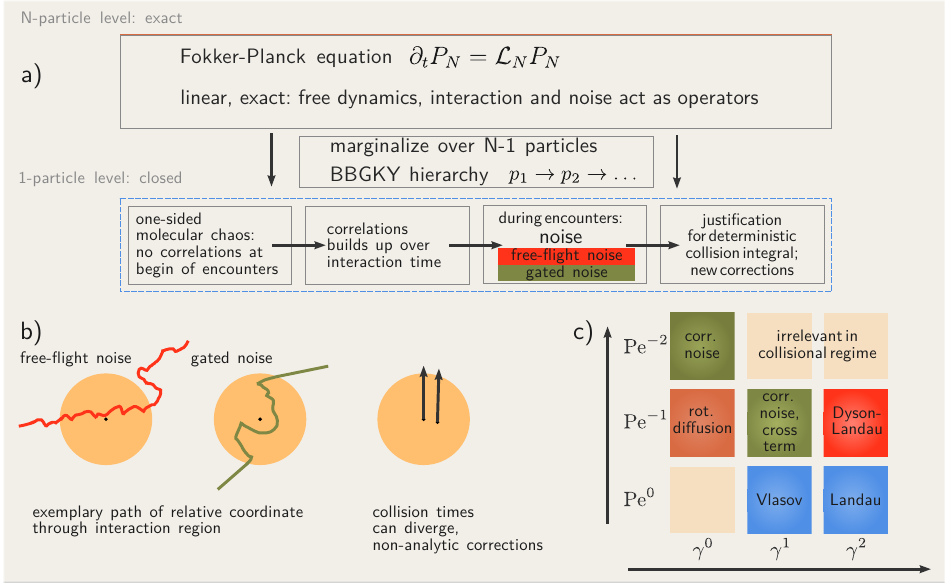}
\caption{Graphical summary of the contributions of this manuscript. a) Closure on the one-particle level introduces interactions between particles which then affect the role of stochasticity in the effective dynamics. We show that using the deterministic collision integral is indeed the correct lowest order and discuss some higher order corrections (panel c)). In particular we are interested in a paradigmatical active matter system (panel b)) of self-propelled particles with alignment interactions for which we discuss two different kinds of noise: a standard omnipresent free-flight noise, that is noisy free-particle dynamics, and one that is gated on contact representing stochastic interactions. This system is not only highly topical, but has several properties making it interesting for this work. Its key property for this work is that it is a collisional system with short-ranged interactions in which the deterministic collision times can be arbitrarily large. As a consequence of this, there are regimes in which any noise is a relevant perturbation leading to non-analytic corrections that we discuss. Panel c) indicates formal orders, the actual contributions for this model are found to be non-analytic.}
\label{fig:overview}
\end{figure}

In this work, we reverse this direction and ask how intrinsic microscopic noise modifies the collision integral itself, cf.~\cref{fig:overview}. The natural case is noise acting on free single-particle dynamics, which we refer to as free-flight noise. What kinetic theories beyond mean-field add is the treatment of interactions as distinct events of finite duration. It is therefore natural to ask what effect noise has during these collisional events. We give a systematic scheme to include weak noise in Landau kinetic theory. The difficulty is that with noise the free two-particle propagator contains a diffusion operator and can no longer be evaluated along deterministic characteristics, while interaction times are not short, and even diverge for nearly parallel particles, so that a short-time expansion is not admissible. We treat this with a Dyson expansion of the propagator to first order in the noise strength, on top of the second order in the interaction strength of Landau theory. We carry this out explicitly for a Kuramoto-Vicsek model of dry, aligning, dilute active matter~\cite{chate2020}. The structure of the model permits a (mostly) analytic evaluation, but the calculation remains intricate, because the time divergences have to be handled.

We further consider a second, physically distinct type of stochasticity which we call {\em gated noise}: noise that acts only while agents interact, as a minimal model of imperfect sensing and response during interactions. Here, the correlations built up during a collision couple directly to the noise, and the divergence of the deterministic collision time becomes relevant. We regularize it with mean passed times obtained from direct simulation of the noisy pair dynamics, which makes the kinetic description semi-analytic.

For noise on the free motion, our main result is a systematic justification of the treatment employed so far~\cite{patelli_2021,boltz2024,soto2024,pinto2025,pinto2026}, in which the noise enters via diffusive terms in the Fokker--Planck equation but not into the collision integral: this is the correct leading order, and the corrections vanish for small noise strengths $\Pe^{-1}\to 0$, where we use the inverse P\'{e}clet number $\Pe^{-1}$ (see below) as a dimensionless quantifier,  even though collision times in the aligning model are unbounded. For noise gated on contact, noise and collisions (which have a strength $\gamma$ and occur with a rate that depends on the typical number of neighbors $M$, see below) do not decouple and the leading correction is of order $\mathcal{O}(\gamma M \Pe^{-1})$ with a logarithmic factor that depends on the noise strength and on magnitude and sign of the coupling. Agent-based simulations show a corresponding contribution odd in $\gamma$, which is absent for noise on the free motion and not captured by regular Vlasov or Landau theory. In the free-flight noise case, this makes kinetic theory with a deterministic collision integral a controlled option for intrinsically noisy agent models, complementary to less model-specific approaches such as the Boltzmann-Ginzburg-Landau approach~\cite{bertin2006,bertin2009,bertin2013,peshkov2014}, in which the noise-renormalized collision integral is modeled directly from symmetry and physical intuition. In the gated noise case, kinetic theory reveals the presence of highly non-intuitive non-analytic terms.

Our results also show the limits of modeling the effect of noise on collisions phenomenologically. Even the minimal model considered here, with bounded interactions and Gaussian white noise, produces contributions that are non-analytic in the noise strength. They stem from nearly parallel pairs, whose collision duration is unbounded, and amount to an orientation-dependent effective noise strength. Symmetry permits such a term, but its coefficient, including its sign, is only fixed by a systematic derivation.

The remainder of this paper is structured as follows. In \cref{sec:landau}, we derive the Dyson-Landau collision integral for free-flight noise and the cross terms for gated noise, the standard derivation of Landau theory and the details of the Dyson expansion are given in \cref{app:landau,app:dyson}. In \cref{sec:daam}, we apply both to the Kuramoto-Vicsek model and compare with agent-based simulations. An evaluation of the collision integrals and angular integrals is given in \cref{app:collintegral,app:gated,app:tsquare,app:deriv}. In \cref{sec:hydro}, we discuss the implications for hydrodynamic theories, and we conclude in \cref{sec:discussion}.

\section{Landau kinetic theory with weak noise} \label{sec:landau}

Before turning to the formalism, we sketch the physical picture, cf.~\cref{fig:overview}. Kinetic theory describes the system by the one-particle distribution function, the probability density to find an agent with a given position and orientation. Its evolution contains a collision integral, which accounts for the effect of encounters between pairs of agents. In mean-field theory, each agent responds to the average state of its neighbors, and encounters do not appear as distinct events. Landau theory goes one step further as it assumes that two agents are uncorrelated when they enter each other's interaction range and computes, to lowest order in the interaction strength, the correlations they build up while they are in contact. For deterministic dynamics, this only requires the straight-line trajectories of the two agents during the encounter. Noise changes this in two ways. The relative trajectory becomes random, so that the duration and geometry of the encounter fluctuate, and the orientations, on which the interaction depends, fluctuate during the encounter. Both effects are small if the encounter is short compared to the time on which the noise acts, and we treat them perturbatively. The exception are nearly parallel agents: their relative velocity is small, so that they stay in contact for a long time, and even weak noise matters for them. These pairs are the origin of the non-analytic contributions found below. If the noise acts only while agents interact (gated noise), it acts precisely during the encounters, and its effect on the correlations appears already at first order in the interaction strength.

In this section, we derive the leading correction of weak noise to the Landau collision integral. The central result is that the noise enters the collision integral only through a commutator, \cref{eq:Jdef}, so that keeping the deterministic collision integral is the correct leading order.

The Landau kinetic equation, originally introduced as a perturbative limit of the Boltzmann equation~\cite{landau1936}, is more accurately described as a systematic perturbative closure for the one-particle distribution function of a weakly interacting many-body system. Its validity regime differs from that of the Boltzmann equation, which relies on a binary collision approximation and is asymptotic in the density: the Landau equation can be a suitable description at arbitrary density, as long as the effective interaction is weak~\cite{boltz2024}. We consider generalized dynamics of $N$ identical agents with single-agent states $\vec j$ (positions and internal degrees of freedom), of Langevin type with short-ranged pair interactions,
\begin{align}
\partial_t {\vec j} &= v_0 \vec{v}_j(\vec j) + \Gamma \sum_{k\neq j} a^R_{jk} \vec{F}_{jk}(\vec j,\vec k) + \op{A}_j\vec{\eta}_j \text{,} \label{eq:langevin}\\
a^R_{jk} &= \Theta(R-\lVert \vec j -\vec k\rVert) \text{,} \label{eq:aij}
\end{align}
with a model-dependent seminorm $\lVert\cdot\rVert$ (typically the Euclidean distance of the positions), bounded interactions $\vec F_{jk}$, and Gaussian white noise $\vec\eta_j$ with unit variance and identical diagonal amplitude matrices $(\op A_j)_{\alpha\beta}=\sqrt{2DD_\alpha}\,\delta_{\alpha\beta}$. The dimensional scale is set by $D$, the relative strengths $D_\alpha$ allow some degrees of freedom to evolve deterministically. The scope and limitations of this model class are discussed in \cref{app:landau}. After rescaling positions by $R$ and times by $R/v_0$, the system is controlled by three non-dimensional parameters: the coupling strength $\gamma = \Gamma R/v_0$, the P\'{e}clet number $\Pe = v_0/(R D)$, and the non-dimensional density $M=V(d) \rho$, where $V(d)$ is the volume of the interaction region and $\rho$ the number density. We consider weakly interacting particles, $\gamma\ll 1$, at arbitrary $M$, and weak noise, $\Pe^{-1}\ll 1$.

Starting from the Fokker--Planck equation equivalent to \cref{eq:langevin}, Landau kinetic theory~\cite{landau1936,balescu,patelli_2021,boltz2024} truncates the BBGKY hierarchy for the reduced distribution functions at order $\mathcal O(\gamma^2)$, using that the pair correlation function $g_2$ is of order $\gamma$ (\cref{app:landau}). This gives
\begin{subequations}
\begin{align}
(\partial_t + \op F_1) p_1(\vec 1,t) &= - \gamma M\,\grad_{\vec 1} \cdot\!\!\int\!\! a_{1\,2} \vec F_{1\,2} \left( p_1(\vec 1,t) p_1(\vec 2,t) {+} g_2(\vec 1,\vec 2,t) \right) \opd\vec 2 \equiv \Jc \text{,} \label{eq:dtp1}\\
(\partial_t + \op F_{1}+\op F_2)\, g_2(\vec 1,\vec 2,t)&= Q(\vec 1,\vec 2,t)= -\gamma\, a_{1\,2}\, \op{L}^{(2)}_{1\,2} \, p_1(\vec 1,t)p_1(\vec 2,t) \text{,} \label{eq:defQ}
\end{align}
\end{subequations}
with the free single-particle Fokker--Planck operator $\op F_i=\op L_i+\Pe^{-1}\op N_i$, consisting of the Liouville operator with $\op L_i p_1=\grad_{\vec i}\cdot(\vec v_i p_1)$ as well as the noise operator $\op N_i=-\sum_\alpha D_\alpha\partial^2_{(\vec i)_\alpha}$, and the pair Liouville operator with $\op L^{(2)}_{1\,2}p_1p_1=\grad_{\vec 1}\cdot(\vec F_{1\,2}p_1p_1)+\grad_{\vec 2}\cdot(\vec F_{2,1}p_1p_1)$. The collision integral $\Jc$ contains a mean-field (Vlasov) contribution from the factorizing part of the pair function and a Landau contribution from pair correlations. Assuming one-sided molecular chaos, i.e.\ uncorrelated particles at the beginning $t_0$ of an interaction, the pair correlations follow as
\begin{align}
g_2(\vec 1,\vec 2,t) &= \int_{t_0}^t \op U_{1\,2}(t',t,\Pe^{\!{-}1})\, Q(\vec 1, \vec 2, t')\,\opd t' \text{,} \label{eq:g2_2}
\end{align}
with the free two-particle propagator $\op U_{1\,2}(t',t,\Pe^{-1})=\exp{-(\op F_1+\op F_2)(t-t')}$.

Without noise, $\op U_{1\,2}$ shifts the arguments along free trajectories, and \cref{eq:g2_2} reduces to the familiar Landau result. With noise, the free propagator contains a diffusion operator and can no longer be evaluated along deterministic characteristics. To the order considered, the one-particle functions evolve freely during a collision, $p_1(\vec 1,t')p_1(\vec 2,t')=\op U_{1\,2}^{-1}(t',t,\Pe^{\!{-}1})\,p_1(\vec 1,t)p_1(\vec 2,t)$, so that
\begin{align}
g_2(\vec 1,\vec 2,t) &= -\gamma\int_{t_0}^t \opd t'\, \op U_{1\,2}(t',t,\Pe^{\!{-}1})\, a_{1\,2}'\op{L}^{(2)}_{1\,2}\, \op U^{-1}_{1\,2}(t',t,\Pe^{\!{-}1})\, p_1(\vec 1,t)p_1(\vec 2,t) \text{,} \label{eq:UAU}
\end{align}
where the prime indicates that the indicator function belongs to the source at time $t'$. Expanding to first order in $\Pe^{-1}$ by means of the Dyson equation (\cref{app:dyson}) yields, for any operator $\op B$,
\begin{align}
\begin{split}
\op U_{1\,2}(t',t,\Pe^{\!{-}1})\,\op B\,\op U^{-1}_{1\,2}(t',t,\Pe^{\!{-}1}) ={}& \op U_{1\,2}(t',t,0)\,\op B\,\op U^{-1}_{1\,2}(t',t,0)\\
&- \Pe^{-1}\!\int_{t'}^t\!\opd s\; \op U_{1\,2}(s,t,0)\big[\op N\bigcomma\, \op U_{1\,2}(t',s,0)\,\op B\,\op U^{-1}_{1\,2}(t',s,0)\big]\,\op U^{-1}_{1\,2}(s,t,0) \text{,}
\end{split} \label{eq:dyson_comm}
\end{align}
with $\op N=\op N_1+\op N_2$. Note that the last line contains a commutator $[ \op C \bigcomma\, \op B] = \op C \op B - \op B \op C$. To avoid confusion, we use brackets $[ \ldots ]$ only in the context of commutators.  This is the analog of first-order time-dependent perturbation theory in the interaction picture of quantum mechanics, with the noise operator $\op N$ as the perturbation~\cite{sakurai} and a similar operator structure is found within projection-operator methods in nonequilibrium statistical mechanics~\cite{zwanzig}. The zeroth order is the deterministic Landau theory, the first order a new Dyson-Landau contribution $\Pe^{\!{-}1} \Jd$,
\begin{align}
\Jc = \Jm + \Jl + \Pe^{\!{-}1} \Jd
\end{align}
with
\begin{subequations}\label{eq:Jdef}
\begin{align}
\Jm&=- \gamma\,M\,\grad_{\vec 1} \cdot\!\!\int\!\!\opd \vec{2}\, a_{1\,2} \vec F_{1\,2}\, p_1(\vec 1,t ) p_1(\vec 2,t) \text{,}\\
\Jl&= \gamma^2 M\,\grad_{\vec 1} \cdot\!\!\int\!\!\opd \vec{2}\, a_{1\,2} \vec F_{1\,2} \int_{t_0}^t \opd t'\, \{ \op U_{1\,2}(t',t,0) a_{1\,2}' \}\{ \op U_{1\,2}(t',t,0) \op{L}^{(2)}_{1\,2} \}\, p_1(\vec 1,t ) p_1(\vec 2,t) \text{,}\\
\Jd &= -\gamma^2 M \,\grad_{\vec 1} \cdot\!\!\int\!\!\opd \vec{2}\, a_{1\,2} \vec F_{1\,2} \int_{t_0}^t \opd t' \int_{t'}^t\opd{s}\; \op{U}(s,t,0) \Big[\op N\bigcomma\, \op{U}(t',s,0)\, a_{1\,2}' \op{L}^{(2)}_{1\,2}\, \op U^{-1}(t',s,0)\Big]\, p_1(\vec 1,s) p_1(\vec 2,s) \text{,}
\end{align}
\end{subequations}
where $p_1(\vec j,s)$ denotes the deterministic free evolution of $p_1(\vec j,t)$ back to time $s$ with $s<t$, which suffices at this order.

The noise thus enters the collision integral only through the commutator of $\op N$ with the propagated source. This commutator vanishes if neither the interaction kernel nor the propagated interaction region depend on the noisy degrees of freedom. The noise acting on the one-particle functions alone, which is already contained in $\op F_1$, drops out. Correspondingly, $\Jd$ decomposes into three contributions, given explicitly in \cref{eq:resultjda} of \cref{app:dyson}: $\Jda$ describes the smearing of the interaction kernel, $\Jdc$ that of the interaction geometry, which yields terms supported on the boundary of the interaction region, and $\Jdb$ is a mixed term.

The expansion is controlled as long as the noise has little effect over the duration $\tau_c$ of a collision. For noise on internal degrees of freedom that set the velocity, the free propagation converts this noise into positional spreading $\propto \Pe^{-1}\tau^3$, so that the relevant parameter is $\Pe^{-1}\tau_c^3$ rather than $\Pe^{-1}$. The expansion is uniform only for bounded collision durations. Within this range, the na\"ive incorporation of noise into Landau theory, by adding diffusive operators and moving from the Liouville to the Fokker--Planck equation while keeping the deterministic collision integral, is correct to leading order.

\subsection{Stochastic interactions, gated noise}

In \cref{eq:langevin}, the noise acts on the free dynamics and is thus an intrinsic property of the agents. In many cases, it is more appropriate to consider the interaction process itself as a source of noise. Generically, an agent interaction in the type of generalized coarse-grained dynamics we are considering here involves some measurement  followed by an adjustment to its outcome. Both steps can be faulty, which gives rise to noise on the level of the model. A minimal version that fits into our framework restricts the noise to particles that are currently interacting, which takes the form of multiplicative noise gated on contact
\begin{align}
\partial_t \vec j &= v_0 \vec{v}_j(\vec j) + \Gamma \sum_{k\neq j} a^R_{jk} \vec{F}_{jk}(\vec j,\vec k) + \sqrt{\sum_{k\neq j} a_{jk}^R}\,\op{A}_j\vec{\eta}_j \label{eq:langevin2} \text{.}
\end{align}
The prefactor $\sqrt{\sum_{k\neq j} a_{jk}^R}$ restricts the noise acting on particle $j$ to times when $j$ is within the interaction range $R$ of at least one other particle. If the noise acts only during interactions, it also acts directly on the pair correlations, and the leading corrections, two cross terms of noise and interaction, appear already at first order in the coupling strength. General systems will combine several noise sources, but we restrict ourselves to this minimal case. We further assume that the noise acts only on degrees of freedom on which $a_{jk}$ does not depend, as in the example below, then the It\^{o} and Stratonovich interpretations coincide~\cite{gardiner}.

A different modeling choice would be to use $\Theta(\sum_{j\neq k} a_{jk}^R)$ as the prefactor. This way noise would either be switched on or off but not increase with the number of neighbors. However, one would then proceed to write $\Theta(\sum_{k\neq j} a_{jk}^R) = 1 - \prod_{k\neq j} (1- a_{jk}^R)$ and find, upon expanding this product and sorting by the number of involved particles, that the two rules coincide on the level of pair interactions.

The terms of the kinetic theory that stem from the deterministic flow $\vec F$ are unchanged. The noise, however, now enters the hierarchy in two places. In the equation for $p_1$, the diffusive term is weighted by the neighbors of particle $1$ and therefore involves the full pair function,
\begin{align}
\begin{split}
(\partial_t + \op L_1) p_1(\vec 1,t) ={}& - \gamma M\,\grad_{\vec 1}\cdot\!\int\!\opd\vec 2\, a_{1\,2} \vec F_{1\,2}\left\{p_1(\vec 1,t)p_1(\vec 2,t) + g_2(\vec 1,\vec 2,t)\right\}\\
&- \Pe^{-1} M\, \op{N}_1 \!\int\!\opd{\vec{2}}\, a_{1\,2} \left\{p_1(\vec{1},t) p_1(\vec{2},t) + g_2(\vec{1},\vec{2},t) \right\} \text{.}
\end{split}\label{eq:bbgkynoise}
\end{align}
In the equation for $g_2$, the noise acting on a pair in contact is a source of correlations by itself, in addition to the interaction,
\begin{align}
(\partial_t + \op L_1+\op L_2)\, g_2(\vec 1,\vec 2,t) &= Q(\vec 1,\vec 2,t) - \Pe^{-1} a_{1\,2}(\op N_1+\op N_2)\, p_1(\vec 1,t)p_1(\vec 2,t) \text{,} \label{eq:g2noise}
\end{align}
with $Q$ from \cref{eq:defQ}. Since it acts only within the interaction range, this source does not factorize. Noise acting on particle $1$ due to contact with a third particle gives rise to the term $-\Pe^{-1}\op N_1\big\{M\!\int\!\opd\vec 3\,a_{1,3}\big[p_1(\vec 1,t)p_1(\vec 3,t)+g_2(\vec 1,\vec 3,t)\big]\big\}\,p_1(\vec 2,t)$, and correspondingly for particle $2$. This is the noise term of \cref{eq:bbgkynoise} multiplied by $p_1(\vec 2,t)$, so that it cancels in the equation for $g_2=p_2-p_1p_1$. The remaining three-particle terms involve $g_2(\vec 1,\vec 2,t)$, $g_2(\vec 2,\vec 3,t)$, or $g_3$ and contribute to the one-particle dynamics only beyond the orders considered here.

We are interested in corrections to first order in $\Pe^{-1}$ to the deterministic Landau theory, which is correct to order $\mathcal{O}(\gamma^2 M)$. Determining all terms of order $\mathcal{O}(\Pe^{-1}\gamma^2 M)$, as for noise on the free dynamics, would involve three-particle terms of the hierarchy. We therefore restrict the analysis to the two leading orders $\mathcal{O}(\Pe^{-1} M)$ and $\mathcal{O}(\Pe^{-1}\gamma M)$. To this order, \cref{eq:g2noise} is solved with the deterministic free propagator as in \cref{eq:g2_2}, and $g_2$ splits into an interaction-induced part of order $\gamma$ and a noise-induced part of order $\Pe^{-1}$. The noise-induced part enters the noise term of \cref{eq:bbgkynoise} only at order $\mathcal{O}(\Pe^{-2}M)$, which we neglect, but it enters the interaction term at order $\mathcal{O}(\gamma M\Pe^{-1})$. With the local neighbor number $M_1=M\int\opd\vec 2\,a_{1\,2}\,p_1(\vec 2,t)$, we find
\begin{align}
\begin{split}
(\partial_t + \op L_1) p_1(\vec 1,t) ={}& \Jm + \Jl - \Pe^{-1} M_1 \op{N}_1 p_1(\vec 1,t)\\
&+ \gamma M \Pe^{-1}\, \op{N}_1\!\int\!\opd{\vec{2}}\, a_{1\,2} \int_{t_0}^t\opd t'\, \{\op U_{1\,2}(t',t,0)\, a_{1\,2}'\} \{\op U_{1\,2}(t',t,0)\, \op{L}^{(2)}_{1\,2}\}\, p_1(\vec 1,t) p_1(\vec 2,t)\\
&+ \gamma M \Pe^{-1}\, \grad_{\vec 1}\cdot\!\int\!\opd{\vec{2}}\, a_{1\,2}\vec F_{1\,2} \int_{t_0}^t\opd t'\, \{\op U_{1\,2}(t',t,0)\, a_{1\,2}'\} \{\op U_{1\,2}(t',t,0)\, (\op N_1+\op N_2)\}\, p_1(\vec 1,t) p_1(\vec 2,t) \text{.}
\end{split}\label{eq:bbgkynoise1}
\end{align}
Here, $\Jm$ and $\Jl$ are the deterministic mean-field and Landau contributions, \cref{eq:Jdef}. The last two lines are the cross terms of noise and interaction: noise acting on the correlations built up by the interaction, and the interaction acting on the correlations built up by the noise. For both, closure at the one-particle level requires only the deterministic propagation of the pair during the interaction, the same input as for Landau theory. The term in the last line has the structure of the Landau collision integral with $\op N_1+\op N_2$ in place of $\op{L}^{(2)}_{1\,2}$ as the collision time always appears multiplied by $\vec F_{1\,2}$. In the second line, this is not the case. For aligning interactions, $\vec F_{1\,2}$ vanishes for parallel particles while their collision time diverges, so that the term in the second line is enhanced logarithmically and dominates to leading logarithmic order, as we show in \cref{sec:gatednoise}.

\section{Application to dry aligning dilute active matter} \label{sec:daam}

In this section, we evaluate these results for a model of aligning self-propelled particles. Its distinctive feature is that nearly parallel particles stay in contact for arbitrarily long times, which renders some of the corrections singular. We apply the general scheme to dry aligning dilute active matter~\cite{chate2020}, for which the deterministic Landau theory is conceptually simple yet practically relevant~\cite{patelli_2021,boltz2024}. The state of each agent separates into its position $\vec r_j$ and internal degrees of freedom $\vec\theta_j$ encoding its orientation, $\vec{j}=(\vec{r}_j,\vec{\theta}_j)^\mathrm{T}$. The free velocity depends only on the orientation, $\vec{v}_j=\vec{v}_j(\vec{\theta}_j)$, and the interactions depend on and act on the orientations only, $\vec{F}_{jk}=\vec{F}_{jk}(\vec{\theta}_j,\vec{\theta}_k)$. Consequently, the orientations have no free dynamics, the velocities are constant during free flight, and the interaction region is determined by the positions alone. Specifically, we consider the model~\cite{peruani2008}
\begin{subequations} \label{eq:model}
\begin{align}
\partial_t \vec r_j &=v_0\, \hat{\vec n}(\theta_j) \text{,}\\
\partial_t \theta_j &= \Gamma \sum_{j\neq i} a_{jk}^{R} \sin(\theta_k-\theta_j) + \sqrt{2D}\, \eta_j \text{,}
\end{align}
\end{subequations}
for $N$ particles in two dimensions with positions $\vec r_j$ and angles $\theta_j$, where $\hat{\vec n}(\theta_j)=(\cos\theta_j,\sin\theta_j)^\mathrm{T}$, $a_{jk}^R = \Theta(R-\lvert \vec r_j - \vec r_k\rvert)$, and $\eta_j$ is Gaussian white noise with $\langle \eta_j \rangle=0$ and $\langle \eta_j(t)\eta_k(t')\rangle = \delta_{jk} \delta(t-t')$. This model has been studied on the Vlasov~\cite{peruani2008}, Landau~\cite{patelli_2021,boltz2024}, and Boltzmann~\cite{ihle2023} level for deterministic dynamics and a non-reciprocal~\cite{fruchart2021,fruchart2026} version has been considered in Ref.~\citenum{mihatsch}. Noise has so far been included on the level of the Fokker--Planck equation, but not in the evaluation of the collision integral.

In the notation of \cref{sec:landau}, the model corresponds to noise on the angles only, with
\begin{align}
\op L_i &= \hat{\vec n}(\theta_i) \cdot \grad_{\vec r_i} \text{,}\qquad
\op N_i = - \partial_{\theta_i}^2 \text{,}\qquad
\op{L}^{(2)}_{1\,2}=(\partial_{\theta_1} -\partial_{\theta_2}) \sin(\theta_2-\theta_1) \text{.}
\end{align}
The norm in the indicator function is the Euclidean distance of the positions, so that
\begin{align}
\op U(t',s,0)\,a'_{1\,2}=\Theta\big(1-\lVert\vec r_{12}-(s-t')\,\vec v_\text{rel}\rVert\big) \label{eq:aprop}
\end{align}
with $\vec r_{12}=\vec r_2-\vec r_1$ and $\vec v_\text{rel}=\hat{\vec n}(\theta_2)-\hat{\vec n}(\theta_1)$, $\lVert\vec v_\text{rel}\rVert=2\lvert\sin(\Delta/2)\rvert$, $\Delta=\theta_2-\theta_1$. In the following, we consider spatially homogeneous one-particle functions, $p_1(\vec 1,t)=p_1(\theta_1,t)$, and use angular Fourier modes,
\begin{align}
\hat{p}_m(t) &= \frac{1}{2\pi} \int_{0}^{2\pi} \exp{-\ii m \theta}\, p_1(\theta,t) \,\opd\theta \text{,}\qquad
p_1(\theta,t) = \sum_{m=-\infty}^\infty \hat{p}_m(t)\, \exp{\ii m \theta} \text{,}
\end{align}
with $\hat{p}_m = \hat{p}_{-m}^*$. For homogeneous $p_1$, the free propagator acts only on the indicator function, the deterministic free evolution of $p_1$ is trivial, and \cref{eq:Jdef} reduces to
\begin{subequations}
\begin{align}
\Jl&= \gamma^2 M\,\partial_{\theta_1} \!\!\int\!\!\opd \vec{2}\, a_{1\,2} \sin(\theta_2-\theta_1) \Big\{\int_{t_0}^t \opd t'\, 1 \Big\} \op{L}^{(2)}_{1\,2}\, p_1(\theta_1,t ) p_1(\theta_2,t) \text{,} \\
\Jd &= -\gamma^2 M \,\partial_{\theta_1}\!\!\int\!\!\opd \vec{2}\, a_{1\,2}\sin(\theta_2-\theta_1) \int_{t_0}^t \opd t' \int_{t'}^t\opd{s}\; \op{U}(s,t,0) \Big[\op{N}\bigcomma\, \{\op{U}(t',s,0)\, a_{1\,2}'\}\, \op{L}^{(2)}_{1\,2}\Big]\, p_1(\theta_1,t) p_1(\theta_2,t) \label{eq:resultjd} \text{.}
\end{align}
\end{subequations}
The expression in braces in $\Jl$, integrated over the interaction region, gives the mean collision duration, which is easily computed either by a geometric averaging argument~\cite{boltz2024} or by considering the limit $t-t_0 \to \infty$ in Fourier space~\cite{balescu,patelli_2021}.

\subsection{Free-flight noise} \label{sec:freenoise}

We evaluate $\Jd$ with the decomposition of \cref{eq:resultjda} in the appendix. The terms $\Jdb$ and $\Jdc$ contain derivatives of the propagated indicator function \cref{eq:aprop} with respect to the angles. These are supported on the boundary of the interaction region, whereas $\Jda$, like the Landau term, receives contributions from the entire interaction region. The explicit propagation and the evaluation of the integrals are given in \cref{app:collintegral}. The collision duration diverges for nearly parallel particles, $\lVert\vec v_\text{rel}\rVert\to 0$. In the Landau term, this is compensated by the interaction kernel $\sin\Delta$, which vanishes in the same limit. This is reflective of physical intuition as two particles that stay close to each other will have similar orientation and, therefore, negligible effective alignment. In the Dyson-Landau term, higher powers of the collision duration enter, and the angular integrals in $\Jdc$ contain a non-integrable singularity at $\Delta=0$. It signals that the first-order Dyson expansion breaks down for nearly parallel pairs, whose collision duration exceeds the time scale $\Pe^{1/3}$ on which the noise becomes relevant. A consistent treatment requires a regularization of the collision duration by the noise itself. This yields a contribution of order $\mathcal O(\gamma^2 M\Pe^{-2/3})$ with the structure $\partial_\theta^2(p_1^2)$, which is non-analytic in the noise strength and whose prefactor depends on the details of the regularization. Since this contribution only modifies the share of nearly parallel pairs in the Landau collision integral, it remains a small fraction of the latter, of order $\mathcal O(\Pe^{-2/3})$, and far below the resolution of the agent-based simulations in \cref{sec:gatednoise}. We therefore do not carry out the regularization for free-flight noise. For gated noise, the corresponding singular term is the leading contribution, and we demonstrate the regularization in detail in \cref{sec:gatednoise}. Here, we give the remaining finite parts of the angular integrals, which we evaluate by contour integration (\cref{app:deriv}). In Fourier space, these finite parts of $\Pe^{-1}\Jd$ split into $\hat{\operatorname{J}}^{\text{D,1}}_\text{coll,n} + \hat{\operatorname{J}}^{\text{D,2}}_\text{coll,n} +\hat{\operatorname{J}}^{\text{D,3}}_\text{coll,n}$, corresponding to $\Jda$, $\Jdb$, and $\Jdc$, respectively, with $\gamma^{-2}M^{-1} \Pe\, \hat{\operatorname{J}}^{\text{D},q}_\text{coll,n} = \sum_k \hat{p}_{n-k}(t) \hat{p}_{k} C_q(n,k)$ and
\begin{align}
C_1(n,k) &= \frac{n\pi^2}{2} \Big\{ 6n\,\delta_{0,k} + \big(((n-2k)^2+2)\sgn(k) + 3(n-2k)\big)\delta_{1,\lvert k\rvert} - 2\big((n-2k)^2+2\big)\sgn(k)\Big\} \\
C_2(n,k) &= -\frac{\pi^2 n}{4} \Big\{16n\,\delta_{0,k} + \big(4(n-2k) + ((n-2k)^2+4)\sgn(k)\big)\delta_{1,\lvert k\rvert} - 4\big((n-2k)(n-6k)+4\big)\sgn(k)\Big\}\\
C_3(n,k) &= - \frac{n \pi^2}{12}\Big\{-2n\, \delta_{0,k} + \big(n-2k+2\sgn(k)\big)\delta_{1,\lvert k\rvert}+4(6k^2-2kn+1)\sgn(k)\Big\}
\end{align}
where we use the convention $\sgn(0)=0$ for the sign function.

The three finite parts are of the same order and partly of opposite sign. The effect of the noise can therefore neither be reduced to a broadening of the interaction kernel, which corresponds to $\Jda$, nor to a modification of the interaction geometry and collision duration, which corresponds to $\Jdc$. Relative to the Landau term, the Dyson-Landau term is suppressed by $\Pe^{-2/3}$ for the singular and by $\Pe^{-1}$ for the regular part. It is irrelevant on the Vlasov level,~\footnote{The Boltzmann equation, in contrast, is correct to all orders in $\gamma$ but asymptotic in $M$. Including noise there would require solving the collision problem for fixed noise realizations, followed by an average over the noise.} which is only correct to $\mathcal O(\gamma M)$.

These corrections are difficult to detect. The mode dynamics is dominated by the mean-field contributions of order $\mathcal{O}(M\gamma)$ and the Landau contributions of order $\mathcal{O}(M\gamma^2)$, which are both less suppressed than the Dyson-Landau corrections. One could envision placing the system at the mean-field transition, $\Pe^{-1}=\pi M \gamma$, which eliminates the bare $\Pe^{-1}$ and $M\gamma$ terms from the polar dynamics. There, however, the Dyson-Landau terms are of order $\mathcal O(M^{5/3}\gamma^{8/3})$ for the singular and $\mathcal{O}(M^2\gamma^3)$ for the regular part, and thus barely separated from contributions beyond the Landau closure, which we do not have access to. Similarly, the onset of polar order $\gamma_c(\Pe^{-1})$ provides no clean test, because the transition in metric polar active matter is driven by density fluctuations~\cite{Martin2018,romanczuk}, which we have neglected. For noise acting on the free motion, the corrections to Landau theory beyond the diffusive term of the Fokker--Planck equation are thus present and non-analytic in the noise strength, but they remain a small fraction of the Landau contribution, even in this model with unbounded collision times.

\subsection{Gated noise} \label{sec:gatednoise}

For noise gated on contact, \cref{eq:langevin2}, the bare noise term is weighted by the local neighbor number, and the leading corrections are the two cross terms of noise and interaction in \cref{eq:bbgkynoise1}, of order $\mathcal O(\gamma M\Pe^{-1})$. For homogeneous one-particle functions, $M_1= M$, the time propagations are trivial, and the mode equations read
\begin{align}
\partial_t \hat{p}_n &= \hat{J}_{\text{coll},n} - n^2 \Pe^{-1} M \hat p_n + \Pe^{-1} \gamma M \big(\hat X_n + \hat{Y}_n\big) \text{,} \label{eq:modesnoise}
\end{align}
where $\hat J_{\text{coll},n}$ contains the deterministic mean-field and Landau terms, and the two noise integrals are
\begin{align}
X&=\partial_{\theta_1} \!\int\!\opd{\vec{2}}\, a_{1\,2}\sin(\theta_2-\theta_1) \Big\{\int_{t_0}^t\opd t'\, 1 \Big\} (\op N_1+\op N_2)\, p_1(\theta_1,t) p_1(\theta_2,t) \text{,}\\
Y&=-\partial_{\theta_1}^2 \!\int\!\opd{\vec{2}}\, a_{1\,2} \Big\{\int_{t_0}^t\opd t'\, 1 \Big\} \op{L}^{(2)}_{1\,2}\, p_1(\theta_1,t) p_1(\theta_2,t) \text{.}
\end{align}
Both only require the deterministic collision duration integrated over the interaction region, $\int\opd\vec r_2\, a_{1\,2}\int\opd t'\,1=4/(3\lvert\sin(\Delta/2)\rvert)$~\cite{boltz2024}. In contrast to free-flight noise, the noisy evolution of the one-particle functions during a collision does not contribute at this order, since the noise acting on a single particle is itself proportional to $M$.

The integral $X$ describes the interaction acting on the pair correlations built up by the noise. It has the structure of the Landau collision integral with $\op N_1+\op N_2$ in place of $\op L^{(2)}_{1\,2}$, and the collision duration always appears multiplied by the interaction kernel $\sin\Delta$. The angular integrals therefore converge, as for the Landau term, and we find (\cref{app:gated})
\begin{align}
\hat X_n = -\frac{32n^2}{3}\sum_k \frac{\big((n-k)^2+k^2\big)\big(4k(n-k)-1\big)}{(4k^2-1)\big(4(n-k)^2-1\big)}\,\hat p_{n-k}(t)\hat p_k(t) \text{,} \label{eq:Xhat}
\end{align}
where we have ``symmetrized'' the coefficients under $k\to n-k$, since only the symmetric part contributes to the sum. For $\lvert k\rvert\gg n$, the coefficients approach $\frac{16}{3}n^2$. The coupling to high modes is thus independent of $k$ and has the form of the Fourier transform of a local term $\propto\partial_\theta^2(p_1^2)$.

The integral $Y$ describes the noise acting on the pair correlations built up by the interaction. Here, the noise operator acts outside of the collision integral, and the collision duration enters without the interaction kernel. In Fourier space, we find (\cref{app:gated})
\begin{align}
\hat Y_n = \frac{2n^2}{3}\sum_k \hat p_{n-k}(t)\hat p_k(t)\Big\{-4 I_9(k) - \frac{32k(n-2k)}{4k^2-1}\Big\} \text{,}\qquad I_9(k)=\int\opd\Delta\,\exp{\ii k\Delta}\,\frac{\cos\Delta}{\lvert\sin(\Delta/2)\rvert} \text{,} \label{eq:Yhat}
\end{align}
and $I_9$ diverges logarithmically at $\Delta=0$. This is the same situation as for free-flight noise in \cref{sec:freenoise}: the deterministic collision duration of nearly parallel pairs diverges, the first-order treatment of the noise breaks down for these pairs, and noise of any strength renders their collision duration finite. Here, however, the singular term is part of the leading coupling of noise and interaction, of order $\mathcal O(\gamma M\Pe^{-1})$, rather than a small correction to the Landau term. We therefore carry out the regularization explicitly. Since the divergence stems from the collision durations alone, it suffices to replace the deterministic durations by the mean passed times of pairs under the noisy pair dynamics. We obtain these from direct simulation of the pair dynamics and use them as input for the kinetic theory, similar in spirit to the use of static structure factors as a closure for steric interactions~\cite{soto2024}. The underlying first-passage problem has been studied for single active particles in a disk~\cite{diTrapani,iyaniwura2025}, but in the opposite limit of strong noise.

The regularization depends on the relative strength of coupling and noise during a collision. The coupling acts on the time scale $\lvert\gamma\rvert^{-1}$, whereas the noise affects the relative motion on the time scale $\Pe^{1/3}$, cf.\ \cref{sec:landau}. We first discuss the noise-dominated case, $\lvert\gamma\rvert\ll\Pe^{-1/3}$, which also covers the parameters of the agent-based simulations below.

\begin{figure}
\includegraphics{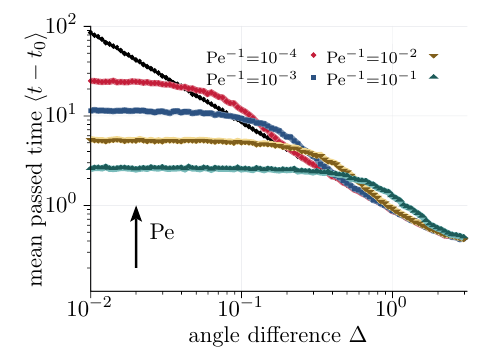}
\caption{Numerically computed mean passed times of particle pairs in the interaction region as a function of the initial angle difference, in log-log scale. For every value of $\Delta$ and $\Pe^{-1}$, $3000$ realizations were sampled. The main effect of the noise is that the diverging passed times of nearly parallel particles, $\tau \sim (\sin(\Delta/2))^{-1}$, are regularized to finite values. For intermediate angle differences, the passed times with noise exceed the deterministic ones: these trajectories mostly follow the deterministic direction, but with a smaller net relative speed due to the noise-induced fluctuations of the relative orientation. This can be seen by noting that the relative velocity at time $t>t_0$ has a Kratky-Porod-like behavior $\langle \vec v_{\text{rel}}(t)\rangle \sim \vec v_{\text{rel}}(t_0) \exp{-(t-t_0)\Pe}$. }
\label{fig:appnoise}
\end{figure}

\subsubsection{Noise dominates, $\lvert \gamma\rvert \ll \Pe^{-1/3}$}

\Cref{fig:appnoise} shows the mean passed times. Their maximum, attained for parallel pairs, scales as $\Pe^{1/3}$. This is directly understable in terms of scaling arguments. In the regime of negligible coupling, the relative motion of nearly parallel pairs is driven by the diffusion of their angle difference, $\langle(\Delta-\Delta_0)^2\rangle=4\Pe^{-1}t$. For $\Delta_0=0$, the relative velocity is $\vec v_\text{rel}\approx\Delta(t)\,\hat{\vec e}_\perp$, with an essentially fixed direction $\hat{\vec e}_\perp$ perpendicular to the mean orientation, so that the relative displacement $\Delta\vec r\approx\hat{\vec e}_\perp\int_0^t\opd t'\,\Delta(t')$ is the integral of a random walk, with $\langle\Delta\vec r^2\rangle=\frac43\Pe^{-1}t^3$. These are the statistics of the transverse deflection of a stiff worm-like chain, with time in place of arc length and $\Pe$ playing the role of the persistence length, and the exit from the interaction region corresponds to the deflection length of a confined semiflexible polymer~\cite{odijk1983,burkhardt1993}. Passage from the interaction region thus takes $\tau\sim \Pe^{1/3}$. The coupling acts on the timescale $\lvert\gamma\rvert^{-1}$, and the comparison of the two scales sets the extent of the noise-dominated regime. From the deterministic result $\langle t-t_0\rangle = \frac{8}{3\pi} \frac{1}{2 \lvert \sin(\Delta/2)\rvert}$ and this scaling analysis, we expect a smooth function $g(u)$ such that
\begin{align}
\frac{8}{3\pi}\langle t-t_0\rangle^{-1} = 2 \lvert \sin(\Delta/2)\rvert + g\big(2 \lvert \sin(\Delta/2)\rvert/\eps\big) \text{,}\qquad \eps\sim \Pe^{-1/3} \text{.} \label{eq:timecorr}
\end{align}

\begin{figure}
\includegraphics{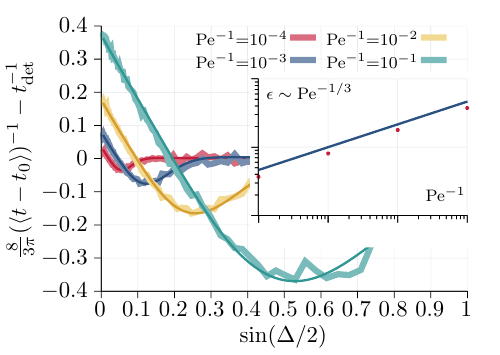}
\caption{Noise-induced correction to the mean passed times, cf.\ \cref{eq:timecorr}. The thin dashed lines are fits of the form $g(u)= \eps \big(\sqrt{u^2+1} -u - a u^2\exp{-b u^2}\big)$ with $u=2 \lvert \sin(\Delta/2)\rvert/\eps$. The latter part was found phenomenologically. This simple form describes the data accurately, particularly for small noise. Inset: the fitted values of $\eps$ (red dots) follow the expected scaling $\eps\sim \Pe^{-1/3}$ as the blue line is the monomial $\Pe^{-1/3}$, so that the prefactor is of order one.}
\label{fig:fits}
\end{figure}

The data are well described by $g(u)= \eps \big(\sqrt{u^2+1} -u - a u^2\exp{-b u^2}\big)$ with $a,b\approx \text{const}$ and $\eps\propto \Pe^{-1/3}$, see \cref{fig:fits}. While the behavior of $g(u)$ at small $u$ follows from the scaling argument, the tail at larger $u$ is a purely phenomenological fit. We use it to obtain a form that is free of statistical noise and, at least in principle, suited for analytic calculations. Numerical quadrature of the mode coupling coefficients based on the measured passed times is a possible alternative that does not require this step. The divergence of $I_9$ only depends on the behavior at small angle differences, where the $a$-dependent part is irrelevant. Replacing $\lvert\sin(\Delta/2)\rvert$ by $\sqrt{\sin^2(\Delta/2)+\delta^2}$ with $\delta=\eps/2$, we find for the regularized integral (\cref{app:deriv})
\begin{align}
\tilde I_9(k) = -4\log\delta + 4\log 4 + 2\big(R_{\lvert k+1\rvert}+R_{\lvert k-1\rvert}\big) + o(\delta^0) \text{,}\qquad R_k = -2\sum_{j=1}^{k}\frac{1}{2j-1} \sim - \log k -2 \log 2- \gamma_{\text{EM}} \text{.} \label{eq:I9reg}
\end{align}
For the asymptotics of $R_k$, we can make use of the standard harmonic numbers $H_k=\sum_{j=1}^k 1/k$ as $R_k = -2 (H_{2k}-H_k/2)$ and then use the well-known Euler-MacLaurin formula $H_k \sim \log k + \gamma_{\text{EM}}$ with the Euler-Mascheroni constant $\gamma_{\text{EM}} \approx 0.577\ldots$.   The leading logarithmic approximation therefore requires $\lvert k\rvert\ll1/\delta$.
The constant $4\log 4$ is specific to the form of the regularization and can be absorbed into the prefactor of $\delta$, whereas the $k$-dependent terms do not depend on the regularization. To leading logarithmic order, with $\delta\propto\Pe^{-1/3}$, the noise integral of \cref{eq:Yhat} becomes
\begin{align}
\hat{Y}_n &= \frac{32n^2}{9} \log(\Pe^{-1}) \sum_k \hat{p}_{n-k}(t)\hat{p}_k(t) + \mathcal{O}(\Pe^0) \text{.}
\end{align}
This is the Fourier transform of $\frac{32}{9}\log(\Pe)\,\partial_\theta^2 \big(p_1^2\big)$, a density-dependent correction to the rotational diffusion that is non-analytic in the noise strength. For aligning interactions, $\gamma>0$, nearly parallel pairs stay in contact longer, so that the noise is enhanced for more populated orientations and for anti-aligning interactions, $\gamma<0$, the effect is reversed.

\subsubsection{Coupling dominates, $\lvert \gamma\rvert \gg \Pe^{-1/3}$}

As apparent from \cref{eq:I9reg}, the leading order of $\tilde I_9$ only depends on the cutoff $\delta$, i.e.\ on the longest passed times, which are set by nearly parallel pairs. For $\gamma\gg \Pe^{-1/3}>0$, we can reduce the analysis to the transversal relative coordinate $y$, which obeys $\ddot{y} \approx -2 \gamma \dot y+ \sqrt{4\Pe^{-1}}\, \xi$. For strong coupling, we can neglect the inertial term and find $\delta \sim \Pe^{-1}/\gamma^2$, corresponding to $\tilde I_9 \simeq -4 \log(\Pe^{-1}/\gamma^2)$. For $\gamma<0$, there is a forbidden zone~\cite{ihle2023}, $\lvert \Delta\rvert <\Delta_c \sim  \lvert\gamma\rvert$, of small angle differences that are not reachable by the deterministic dynamics. This leads to $\delta\sim \lvert\gamma\rvert$ and $\tilde I_9 \simeq -4 \log\lvert\gamma\rvert$. The same scaling follows from noting that the deterministic turning takes the time $1/\lvert\gamma\rvert$, as does the diffusive process required to reach angle differences large enough for the anti-alignment to dominate. These asymptotics are consistent in the sense that they match the noise-dominated result for $\lvert \gamma\rvert \sim\Pe^{-1/3}$.

\begin{figure}
\includegraphics[]{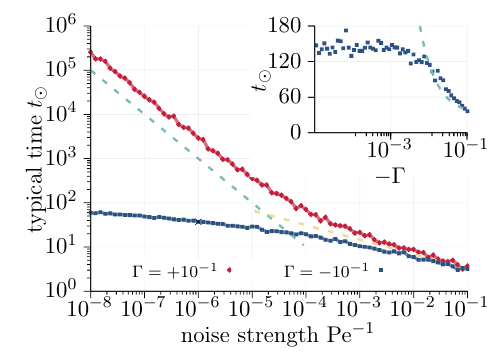}
\caption{Typical passed times. To avoid the computational cost of mean times, we consider a single realization with particles starting at the same position and with identical orientations. The actual mean-passed time would have required to consider all possible end points within the end circle as well as all realizations of the noise. As the times are long, sample-to-sample fluctuations are relatively small, so that this is a good indicator for the scaling of the passed time. The analytically predicted scaling behaviors (dashed lines) are reproduced: We see a decay like $1/\Pe^{-1}$ for aligning particles and small noise (coupling dominates, teal dashed line), whereas both aligning and anti-aligning coupling shows a shallow $1/\Pe^{-1/3}$ decay at large (dominating) noise (amber dashed line). In the anti-aligning case, the passed time depends only weakly on the noise strength in the coupling-dominated regime (small noise). Instead it depends inversely on the coupling strength (inset). The location of this cut along the coupling axis is indicated by a small black cross in the main plot.}
\label{fig:maxtimes}
\end{figure}

\subsubsection{Crossover between the asymptotic regimes}

Combining the asymptotic regimes, we use \cref{eq:I9reg} with the interpolation
\begin{align}
\delta(\gamma,\Pe^{-1}) = \begin{cases}\sqrt{\Pe^{-2/3}+\gamma^2} & \gamma \leq 0\\
	\dfrac{\Pe^{-1}}{\gamma^2+\Pe^{-2/3}} & \gamma > 0
\end{cases} \text{.}
\end{align}
To leading logarithmic order, the noise integral then reads
\begin{align}
\hat{Y}_n &= \frac{32n^2}{3} \log\delta(\gamma,\Pe^{-1}) \sum_k \hat{p}_{n-k}(t)\hat{p}_k(t) + \mathcal{O}(\delta^0) \text{,}
\end{align}
and the full expression follows from \cref{eq:Yhat,eq:I9reg}.

\subsubsection{Comparison with agent-based simulations}

\begin{figure}
\includegraphics{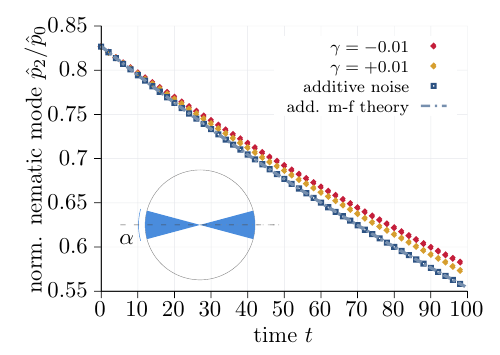}
\caption{Normalized nematic mode $\hat{p}_2/\hat{p}_0$ against time for $M=0.1$, $\Pe^{-1}=0.01$, and $\gamma=\pm 0.01$, from agent-based simulations (symbols) with a stochastic Heun scheme and time step $\Delta t=0.005$. The used values put this simulation in the noise-dominated regime. The normalization allows a direct comparison with the integrated mode equations. For additive noise of strength $\Pe^{-1}M$, mean-field theory (line) agrees with the simulations to high precision. For gated noise, the relaxation depends on the sign of $\gamma$, a signature of the cross terms $X$ and $Y$, and both curves relax more slowly than for additive noise, see text. Inset: initial condition. The particles are spatially homogeneous, with orientations drawn from a double cone of opening angle $\alpha=\pi/6$ around the $x$-axis. This excites only the real parts of the even modes, and the polar mode vanishes throughout.}
\label{fig:stochmodes}
\end{figure}

Our aim here is not a quantitative prediction for this toy model, but to show that the cross terms of noise and interaction are present in the agent-based dynamics and can be identified. The collision-based approach here requires a disordered state. The ordered state has an almost fixed interaction topology~\cite{mora2016} and there is no sensible definition of collisions anymore. In \cref{fig:stochmodes}, we therefore initialize the particles spatially homogeneous with orientations drawn from a double cone, which excites only the even modes. The polar mode, to which all mean-field interaction terms couple, then vanishes throughout, and the nematic mode is the slowest remaining mode. We use $M=0.1$, $\Pe^{-1}=0.01$, and $\lvert\gamma\rvert=0.01$, which lies deep in the noise-dominated regime, $\lvert\gamma\rvert\ll\Pe^{-1/3}$.  We do note that we are over the mean-field threshold for polar order and expect to see strong mean-field coupling in the limit of long times, we find numerically that the polar mode is sufficiently small throughout our relatively short simulation window.

For additive noise with the same mean strength, $\Pe^{-1}M$, mean-field theory agrees with the simulations to high precision. For gated noise, the relaxation depends on the sign of $\gamma$. This dependence is not captured by Vlasov or Landau theory, even if the bare diffusive term is included: the mean-field terms vanish for this initial condition, and the Landau term is even in $\gamma$. To the order considered, the only contributions odd in $\gamma$ are the cross terms $X$ and $Y$ as higher orders in the interaction are suppressed by $\gamma^2$, and no such split occurs for additive noise. The theory also predicts the sign of the split: the coefficients of $\hat p_0\hat p_2$ in both $\hat X_2$ and $\hat Y_2$ are negative, so that the nematic mode relaxes faster for aligning than for anti-aligning interactions, as observed.

In addition, both curves for gated noise relax more slowly than the one for additive noise. This difference is even in $\gamma$ and cannot stem from the Landau term, which is the same for both types of noise. It is consistent with the contributions of order $\mathcal O(\Pe^{-2}M)$ that we have neglected, in which the pair correlations induced by the noise feed back into the noise term. At $\lvert\gamma\rvert=\Pe^{-1}$, they are of the same order as $X$ and $Y$, and they are enhanced by long collision durations in the same way as $Y$. Physically, the number of neighbors fluctuates, so that the gated noise acting on a given particle is intermittent, which slows down the decay of the modes compared to noise of the same mean strength. We cannot rule out higher order terms in $\gamma$, such as $\mathcal{O}(M\gamma^3)$ terms, in principle, but highly suspect that these are not relevant on the relatively short timescales considered here.

A quantitative comparison would require the regular parts of the angular integrals in addition to the leading logarithm, as well as the even contributions of order $\mathcal O(\Pe^{-2}M)$. The former depend on the passed times beyond their small-angle asymptotics, cf.\ \cref{fig:fits}, and are accessible by numerical quadrature with measured passed times as input, whereas the latter require the expansion to second order in $\Pe^{-1}$. Both are possible, but at considerable cost, and we do not pursue them here. The purpose of this example is to show that the scheme applies to collision problems with genuinely singular collision durations and how the resulting singularities can be handled, rather than to provide quantitative predictions for this particular model.

\subsection{Hydrodynamic implications} \label{sec:hydro}

Hydrodynamic theories of aligning active matter are either derived from microscopic models, mostly on the mean-field level, or postulated from symmetry, with coefficients that are possibly obtained from a phenomenological collision rule, as in the Boltzmann-Ginzburg-Landau approach~\cite{bertin2006,bertin2009,bertin2013,peshkov2014}. The singular contributions found above illustrate the gap between these routes. They are permitted by symmetry, but neither mean-field nor Landau theory with a deterministic collision integral contain them. Their coefficients are non-analytic in the model parameters, depend on whether the noise acts during free flight or during interactions, and, for gated noise, change sign with the alignment. Such coefficients can neither be guessed nor be obtained from an instantaneous noisy collision rule, which has no collision duration. They only follow from a systematic treatment of the noise during collisions. In the following, we show that these contributions share a common structure, $\partial_\theta^2(p_1^2)$, and amount to an orientation-dependent effective noise strength.

The common structure follows from the fact that the singular contributions stem from nearly parallel pairs. For gated noise, the regularized collision duration in $Y$ localizes the angular integral at $\Delta=0$ to leading logarithmic order, cf.\ \cref{app:deriv},
\begin{align}
\int\!\opd\theta_2\, \frac{4\,G(\theta_1,\theta_2)}{3\sqrt{\sin^2(\Delta/2)+\delta^2}} = \frac{16}{3}\abs{\log\delta}\, G(\theta_1,\theta_1) + \mathcal O(\delta^0)
\end{align}
for any smooth function $G$. For $\theta_2=\theta_1$, the pair source reduces to a local expression,
\begin{align}
\op L^{(2)}_{1\,2}\,p_1(\theta_1,t)p_1(\theta_2,t)\big\rvert_{\theta_2=\theta_1} = -2F'(0)\,p_1(\theta_1,t)^2 \text{,}
\end{align}
where $F(\Delta)$ is the alignment torque, $F(\Delta)=\sin\Delta$ and $F'(0)=1$ for the model \cref{eq:model}. Hence, $Y = \frac{32}{3}F'(0)\abs{\log\delta}\,\partial_\theta^2\big(p_1^2\big) + \mathcal O(\delta^0)$, and the noise terms in \cref{eq:bbgkynoise1} combine to
\begin{align}
\Pe^{-1}M\,\partial_\theta^2\Big\{\Big(1+\frac{32}{3}\,\gamma F'(0)\abs{\log\delta}\,p_1(\theta,t)\Big)\,p_1(\theta,t)\Big\} \text{,} \label{eq:effnoise}
\end{align}
up to the regular contributions of order $\mathcal O(\gamma M\Pe^{-1})$. The effective noise strength thus depends on the orientation. Aligning interactions build up positive correlations between nearly parallel particles, which remain in contact for long times. A particle with a more populated orientation therefore has more neighbors on average and experiences more of the gated noise. For anti-aligning interactions, the effect is reversed. For free-flight noise, the singular part of $\Jdc$, \cref{sec:freenoise}, stems from the same nearly parallel pairs and has the same structure, with a coefficient of order $\mathcal O(\gamma^2M\Pe^{-2/3})$ that depends on the details of the regularization.

This structure is generic. A contribution localized at $\Delta=0$ is local in orientation and quadratic in $p_1$. Conservation of the particle number requires the form $\partial_\theta J$, and for achiral interactions, a current $J\propto p_1^2$ is excluded by the symmetry $\theta\to-\theta$, so that $J\propto\partial_\theta(p_1^2)$ is the lowest-order form.

In terms of the modes, \cref{eq:effnoise} contributes $\propto n^2\sum_k\hat p_{n-k}\hat p_k$. For the polar mode, the term $2\hat p_0\hat p_1$ is an orientation-dependent correction to the rotational diffusion, which shifts the order-disorder threshold at relative order $\mathcal O(\gamma\abs{\log\delta})$. The term $2\hat p_{-1}\hat p_2$ is a coupling of the polar and nematic modes, which, after adiabatic elimination of the nematic mode, controls the saturation of polar order. Corresponding terms enter the equations for the higher modes. Spatial gradients, which we have not considered here, will give rise to further terms of this type.

\section{Discussion} \label{sec:discussion}

In summary, we introduced a systematic scheme to extend the Landau-theory as the most generally applicable kinetic theory beyond the mean-field approach to stochastic dynamics by way of a Dyson expansion of the noisy free propagator. The noise enters the collision integral in the form of the commutator of noise operator and the time-propagated correlation source. This vanishes if neither the interaction kernel nor the interaction geometry depend on the noisy degrees of freedom. Our most important result is that the na\"ive treatment, using a deterministic collision integral along-side the standard diffusion term from the Fokker--Planck equation, is indeed the correct leading order. We are therefore able to establish a solid theoretical foundation to quantitatively working approaches in the literature~\cite{patelli_2021,boltz2024,soto2024,pinto2025,pinto2026}. However, this is found to be effectively an expansion in the collision times and, thus, the expansion is analytic only if these times are bounded. In the standard workhorse model of active matter, self-propelled particles with alignment interactions, this is not the case and non-analytic terms  $\mathcal O(\gamma^2M\Pe^{-2/3})$ emerge. Both these and the regular finite parts of order $\mathcal O (\gamma^2M\Pe^{-1})$ are effectively too small to be resolved in agent-based simulations. This is, however, likely model-specific and we can imagine these higher orders to be relevant in specific applications. In terms of complexity reduction, we find that while we can identify individual contributions to the new noise-correction of the collision integral with somewhat clear interpretations, these are all of similar order and partially cancel each other. Thus, there is no way to reduce the noise effect just to a broadening of the interaction or subsume it into effective collision times.
We also considered a noise that only acts during interactions, which we refer to as gated noise in contrast to the free-flight noise of the first part. Here, we find that the restriction of the noise itself and the interactions induce pair correlations that on the one-particle level lead to two cross-terms, a noise correction to the interaction and an interaction correction to the noise. While these appear similar, the latter is again non-analytic. Here, we perform an explicit analysis of the regularization the singular perturbation by the noise induces and can deduce logarithmic corrections. We find that both new terms are odd in the coupling strengths, predicting a sign-dependence in the relaxation behavior even if the mean-field channel is suppressed by choice of initial conditions. This is also found in agent-based simulations. A quantitative comparison is impeded by contributions of order $\mathcal O(\Pe^{-2}M)$, which are also enhanced logarithmically, but even in $\gamma$, so that they do not affect the sign dependence. We did not calculate them, as we aimed at general insight rather than quantitative agreement. Furthermore, we mostly limited the analysis on spatially homogeneous states to reduce the number of terms and since we systematically build on collisional interactions, favoring the unordered phase for our methodology.

A larger implication of the present work is a message of caution. We studied the arguably simplest model in this context imaginable with metric bounded finite-range interactions, white Gaussian noise, highly favorable structure of the degrees of freedom and still end up with formally singular, non-analytic terms. These are highly non-intuitive, not obvious from the equations of motion and not present in most symmetry arguments. Phenomenological theory is therefore likely to be unaware of these pitfalls. Effective modelling of noise-renormalized collisional rules based on symmetry and intuition are highly at risk to miss physical effects induced by these non-trivial terms as they will never incorporate them. Also, this work serves as a stark reminder of the importance of collision times~\cite{ihle2023} in kinetic approaches to statistical physics. Effective instantaneous rules would never reproduce the non-analytic terms. As such, similar findings are to be expected at least within the wider class of Vicsek-like models that have identical propulsion speeds. As we have previously shown in dealing with similar non-analytic behavior within Boltzmann theory~\cite{ihle2023}, they can be lifted by introducing a regularizing quenched speed disorder. We see that the source of the noise is crucially important, the relevance of the noise depends on when it acts (in our case free-flight or gated). The relevance of this modelling decision has to our knowledge not been appreciated so far.

For this work, our focus was on general statements and an understanding of the structure of the theory rather than making quantitative predictions. For those, we argue for a semi-analytic approach: Use empirical data for the collisional times of noisy interactions as additional input for the kinetic theory. In a way this can be considered as the dynamical analog to using the static structure factor as closure for steric interactions~\cite{soto2024}. An alternative route is also inspired by the treatment of steric interactions~\cite{bruna2022}, it may be possible to employ matched asymptotics between interaction regions with diverging times, in our case almost parallel pairs, and the rest.

Several extensions suggest themselves as subjects of future work. First, we have restricted the analysis to spatially homogeneous states. With spatial gradients, the free propagator also acts on the positions, which gives rise to couplings of noise and gradients. These are required for hydrodynamic equations in the proper sense and, e.g., for the description of density band formation. Second, the scheme only requires that the generator of the free dynamics decomposes into a deterministic part and a small perturbation. It therefore applies to other sources of stochasticity as well, such as tumbling, fluctuations of speed or chirality, or colored noise by means of auxiliary variables. For chiral particles~\cite{markovich2019,liebchen}, symmetry additionally permits a drift term $\propto\partial_\theta(p_1^2)$ the occurrence of which would be interesting. Third, in experimental realizations of active matter, such as vibrated granular particles or robotic swarms, noise is often generated by contacts, so that gated noise is a natural model. Since steric repulsion bounds the collision durations there, we expect the cross terms of noise and interaction, odd in the coupling, at order $\mathcal O(\gamma M\Pe^{-1})$, but without logarithmic enhancement. Finally, the scheme carries over to non-additive interactions~\cite{bera}, since the noise enters only via the free propagator.

\acknowledgments

H-H.~B is funded by the Deutsche Forschungsgemeinschaft (DFG, German Research Foundation) -- 587108963.

\appendix

\section{Landau kinetic theory} \label{app:landau}

In this appendix, we summarize the standard derivation of \cref{eq:dtp1,eq:defQ,eq:g2_2} and the assumptions on the model class \cref{eq:langevin}.See Refs.~\citenum{balescu,kreuzer,boltz2024} for details.

\subsection{Model class}

We only consider bounded interactions. This is an innate limitation of the Landau approach, which treats the interaction perturbatively. Singular interactions, such as hard-core repulsion, cannot be treated perturbatively and require the full Boltzmann theory~\cite{balescu,dorfman2021,ihle2023}. The indicator function \cref{eq:aij} restricts us to interactions with a fixed range $R$. This excludes two notable classes of models: long-ranged interactions, such as electrostatic forces, which are treated within kinetic theory by means of the Lenard--Balescu equation~\cite{balescu}, and metric-free models~\cite{ballerini2008,ginelli2010metric,chou2012,degond2019}, where adjacency is determined topologically. While the latter feature interactions that fit into a collisional paradigm, they are outside of the scope of this work as determining the beginning and end of interactions is more involved~\cite{chou2012,kuersten2024}. Similarly, we do not explicitly consider non-additive or genuinely multiparticle interactions, but the scheme carries over~\cite{bera}, since the noise enters only via the free propagator.

The scale parameters $v_0$ and $\Gamma$ allow for dimensional analysis, with the understanding that the remaining fields $\vec{v}_j$, $\vec{F}_{jk}$ are of order one. The noise strength is parametrized analogously: $D$ sets the overall strength, and the dimensionless numbers $D_\alpha$ are relative strengths whose purpose is mainly to allow for some degrees of freedom to evolve deterministically ($D_\nu=0$ for some $\nu$). A single overall noise strength only simplifies the presentation and is not a fundamental limitation of the method. The noise is Gaussian and white,
\begin{align}
\langle \vec \eta_j \rangle &=0,\qquad \qquad
\langle (\vec  \eta_j)_\alpha(t) (\vec  \eta_k)_\beta(t')\rangle = \delta_{jk}\delta_{\alpha\beta}\delta(t-t') \text{.}
\end{align}
The definition of $\Pe$ assumes noise on dimensionless internal degrees of freedom. For positional noise, the corresponding $D_\alpha$ is measured in units of $R^2$, which leaves the rescaled equations unchanged.

\subsection{BBGKY hierarchy}

After rescaling units such that $R=v_0=1$, standard stochastic calculus~\cite{gardiner,risken} gives the Fokker--Planck equation for the full $N$-particle distribution function $P_N$ equivalent to the Langevin equation, \cref{eq:langevin},
\begin{align}
	\partial_t P_N(\{\vec j\}) &=  -\sum_{j=1}^N \left\{\grad_{\vec j} \cdot \left(\vec{v}_j(\vec j) P_N\right) + \gamma\grad_{\vec j} \cdot \left( \sum_{k=1,j\neq i}^N a_{jk} \vec F_{jk}(\vec j,\vec k) P_N\right) - \Pe^{\!{-}1}\tilde{\Delta}_{\vec j} P_N \right\} \text{,}
\end{align}
with $\tilde{\Delta}_{\vec j} = \sum_\alpha D_\alpha \partial^2_{j_\alpha}$ being the appropriately weighted Laplacian with respect to the degrees of freedom that belong to particle $j$. Here and in the following, the omitted arguments of $P_N$ are the same on the right as on the left to increase legibility. As we consider all particles to be identical, most relevant quantities, in particular any order parameter, should be reducible to integrals over a {\em one-particle} distribution. Thus, we introduce reduced $k$-particle distribution functions by integrating out all degrees of freedom but those of the first $k$ particles,
\begin{align}
p_k &= \frac{N!}{(N-k)!} \left(\frac{1}{M}\right)^k \int P_N \prod_{j=k+1}^{N}\opd\vec{j} \text{.}
\end{align}
The prefactor $(1/M)^k$ is chosen such that all $p_k$ are of order $\mathcal{O}(1)$ for a uniform distribution $P_N$. In the free problem, $\gamma\equiv 0$, all particles are uncorrelated (if they were initially uncorrelated) and the $k$-particle functions factorize, $p_{k,\text{free}} = \prod_{i=1}^k p_1(\vec i,t)$. Thus, an {\em Ursell} (or {\em Mayer} or {\em cluster}) expansion~\cite{Ursell_1927,mayer} in terms of {\em reduced correlation functions} $g_k$ is used. The correlation function $g_k$ contains the information on the $k$-particle level $p_k$ that is not contained in the lower order $p_j$ and $g_j$ with $j<k$. In particular, the first three correlation functions are
\begin{align}
\label{eq:ursell}
\begin{split}
p_1(\vec 1, t) &= g_1(\vec 1, t)\\
p_2(\vec 1,\vec 2, t) &= p_1(\vec 1, t) p_1(\vec 2, t) + g_2(\vec 1,\vec 2, t)\\
p_3(\vec 1,\vec 2, \vec 3, t)&= p_1(\vec 1, t) p_1(\vec 2, t) p_1(\vec 3, t) {+} p_1(\vec 1, t) g_2 (\vec 2,\vec 3, t) {+} \ldots {+} g_3(\vec 1,\vec 2,\vec 3, t)
\end{split}
\end{align}
where the ellipsis corresponds to two index-permuted terms of the form $p_1 g_2$. The marginalization leads to the textbook~\cite{balescu,kreuzer,cercignani1993,*villani2002,*cercignani2012,*dorfman2021} Born-Bogoliubov-Green-Kirkwood-Yvon (BBGKY) hierarchy~\cite{yvon,*bogolioubov,*born,*kirkwood}, in which the evolution of the $k$-particle functions depends on the $(k+1)$-particle function. In Landau kinetic theory, a systematic expansion in $\gamma$ is performed~\cite{landau1936,balescu,patelli_2021,boltz2024}. One consequence is that the correlation functions scale as $g_k\sim \gamma^{k-1}$, which induces a natural way of closing the hierarchy by terminating at a fixed order of $\gamma$. To order $\mathcal{O}(\gamma^2)$, this yields \cref{eq:dtp1,eq:defQ}.

In \cref{eq:defQ}, we have separated the indicator function $a_{1\,2}$ from the pair Liouville operator $\op{L}^{(2)}_{1\,2}$, as this is helpful for the evaluation. This presupposes that $\vec F_{1\,2}$ has no components along the degrees of freedom on which $a_{1\,2}$ depends, as is the case for the aligning interactions considered in \cref{sec:daam}. Otherwise, an additional term $\propto \vec F_{1\,2}\cdot\grad a_{1\,2}$ appears, supported on the boundary of the interaction region. For reciprocal interactions, $\vec F_{2\,1}=-\vec F_{1\,2}$, the pair operator simplifies to $\op L^{(2)}_{1\,2}=(\grad_{\vec 1}-\grad_{\vec 2})\cdot\vec F_{1\,2}$.

\subsection{Pair correlations}

\Cref{eq:defQ} allows for a direct formal enslaved solution, i.e.\ determining $g_2$ as a functional $g_2[p_1]$ of the one-particle function, by variation of constants (Duhamel's principle),
\begin{align}
	g_2(\vec 1,\vec 2,t) &= \op U_{1\,2}(t_0,t,\Pe^{\!{-}1})g_2(\vec 1,\vec 2,t_0) {+} \op U_{1\,2}(t_0,t,\Pe^{\!{-}1})\int_{t_0}^t  \!\!\!\op U_{1\,2}^{-1}(t_0,t',\Pe^{\!{-}1}) Q(\vec 1, \vec 2,t')\,\opd t' \text{,} \label{eq:g2duhamel}
\end{align}
with the free two-particle propagator
\begin{align}
	\op U_{1\,2}(t_0,t,\Pe^{\!{-}1}) = \exp{(-\op F_1-\op F_2)(t-t_0)} = \exp{-\op F_1(t-t_0)}\exp{-\op F_2(t-t_0)}\text{,}
\end{align}
where we have used that the free propagators of different particles commute. The error made by any approximation of the time evolution generally grows with the propagation time, so that a reasonable choice of $t_0$ is required, together with the value of $g_2$ at that time. In the weakly interacting situation considered here, there will generally be large time gaps between subsequent collisions of the same pair of particles. This leads naturally to the assumption of {\em one-sided} molecular chaos: particles are taken to be uncorrelated at the beginning of an interaction, which for finite $R$ is clearly defined~\cite{kreuzer,balescu,boltz2024}. This is to be contrasted with the more common assumption of molecular chaos, i.e.\ factorization of the multiparticle probability functions throughout. Disregarding correlations entirely leads to mean-field (or Vlasov) dynamics, which is correct to order $\mathcal{O}(\gamma M)$. Identifying $t_0$ as the beginning of the interaction, one-sided molecular chaos translates to $g_2(\vec 1,\vec 2,t_0)=0$. With the propagator property $\op U_{1\,2}(t_0,t,\Pe^{\!{-}1}) \op U_{1\,2}^{-1}(t_0,t',\Pe^{\!{-}1})=\op U_{1\,2}(t',t,\Pe^{\!{-}1})$, this gives \cref{eq:g2_2}.

\section{Dyson expansion of the collision integral} \label{app:dyson}

\subsection{First-order Dyson expansion}

In the deterministic case, $\Pe^{-1}=0$, the free propagator shifts the arguments along the free trajectories, found from integrating $\partial_t \vec i = \vec{v}_i$, which yields the well-known property $\op U(t',t) A(t')B(t')=A(t)B(t)$ of deterministic propagators~\cite{zwanzig}. For finite $\Pe$, this property is lost. Instead, we expand the propagator to first order in the inverse P\'{e}clet number,
\begin{align}
\op U_{1\,2}(t',t,\Pe^{\!{-}1}) &= \op U_{1\,2}(t',t,0) - \Pe^{-1} \int_{t'}^t\opd{s}\, \op{U}(s,t,0) \op{N} \op{U}(t',s,0) \text{,} \label{eq:firstW}
\end{align}
which reduces the evaluation to the deterministic free propagator. To see this, note that $\partial_t \op U_{1\,2} (t_0,t,\Pe^{-1}) = - (\op F_1+\op F_2) \op U_{1\,2}(t_0,t,\Pe^{-1})$. With $\op N=\op N_1+\op N_2$, this yields
\begin{align}
\frac{\opd{}}{\opd s} \{ \op U_{1\,2} (s,t,0) \op U_{1\,2} (t',s,\Pe^{-1}) \} = - \Pe^{-1} \op U_{1\,2} (s,t,0) \op N \op U_{1\,2} (t',s,\Pe^{-1}) \text{.}
\end{align}
Integrating over $s$ from $t'$ to $t$ gives the exact Dyson equation~\cite{dyson1949,feynmann1951}. Replacing the full propagator on its right-hand side by the deterministic one yields \cref{eq:firstW} to first order in $\Pe^{-1}$. Correspondingly, the inverse propagator is
\begin{align}
\op U^{-1}_{1\,2}(t',t,\Pe^{\!{-}1}) &= \op U^{-1}_{1\,2}(t',t,0) + \Pe^{-1} \int_{t'}^t\opd{s}\, \op{U}^{-1}(t',s,0)\, \op{N}\, \op{U}^{-1}(s,t,0) \text{,}
\end{align}
where we have used $\op U(t',t,0)=\op U(s,t,0)\op U(t',s,0)$. Inserting both expansions into $\op U_{1\,2}(t',t,\Pe^{-1})\,A\,\op U^{-1}_{1\,2}(t',t,\Pe^{-1})$ and collecting the terms of first order gives \cref{eq:dyson_comm}.

\subsection{Explicit form}

Applying \cref{eq:dyson_comm} to \cref{eq:UAU} and using the deterministic back-evolution $\op U^{-1}_{1\,2}(s,t,0)\,p_1(\vec 1,t)p_1(\vec 2,t)=p_1(\vec 1,s)p_1(\vec 2,s)$, we find
\begin{align}
	\begin{split}
	g_2(\vec 1,\vec 2,t) ={}& -\gamma\,\int_{t_0}^t\opd t'\, \{\op U_{1\,2}(t',t,0) a_{1\,2}'\} \{\op U_{1\,2}(t',t,0) \op{L}^{(2)}_{1\,2}\} p_1(\vec 1,t) p_1(\vec 2,t) \\
	&+\gamma\,\Pe^{-1} \int_{t_0}^t\opd t'\,\int_{t'}^t\opd{s}\, \op{U}(s,t,0) \op{N} \{\op{U}(t',s,0) a_{1\,2}'\} \{\op{U}(t',s,0) \op{L}^{(2)}_{1\,2}\} p_1(\vec 1,s) p_1(\vec 2,s)\\
	&-\gamma\,\Pe^{-1} \int_{t_0}^t\opd t'\,\int_{t'}^t\opd{s}\, \op{U}(s,t,0) \{\op{U}(t',s,0) a_{1\,2}'\} \{\op{U}(t',s,0) \op{L}^{(2)}_{1\,2}\}\, \op{N}\, p_1(\vec 1,s) p_1(\vec 2,s) \text{.}
	\end{split}\label{eq:g2}
\end{align}
The last line stems from expressing $p_1(t')$ through $p_1(t)$ with the noisy free propagator. It removes the free diffusion of the one-particle functions during the collision, which is already part of their evolution from $t'$ to $t$.

\subsection{Decomposition}

Conceptually, $t_0$ is the beginning of the collision and $t>t_0$ cannot exceed the end time of the collision. Therefore, everything is evaluated {\em during} collisions, and one would want to set $\{ \op U_{1\,2}(t',t,0) a_{1\,2}' \} = \{\op{U}(t',s,0) a_{1\,2}'\} =1$. The first is unproblematic, but the latter is not, as $\op N$ acts on the propagated indicator function and its derivatives are relevant. Some care is needed in the subsequent integrations, as the derivatives of the Heaviside function are supported on the boundaries of the time integration. With respect to the phase-space coordinate $\vec{2}$, however, these are interior points. The derivatives yield $\delta$ and $\delta'$ terms supported on the boundary of the interaction region. Also noting that the interaction and the noise are not explicitly time-dependent, we find
	\begin{subequations}\label{eq:resultjda}
		\begin{align}
	\Jl&=  \gamma^2 M\,\grad_{\vec 1} \cdot\!\!\int\!\!\opd \vec{2}\, a_{1\,2} \vec F_{1\,2} \int_{t_0}^t \opd t'\, \{\op{U}(t',t,0) \op{L}^{(2)}_{1\,2}\}   p_1(\vec 1,t ) p_1(\vec 2,t)   \\
	\Jd &= -\gamma^2 M \,\grad_{\vec 1} \cdot\!\!\int\!\!\opd \vec{2}\, a_{1\,2} \vec F_{1\,2}  \int_{t_0}^t \opd t'\,  \int_{t'}^t\opd{s}\, \{\Jda-2\Jdb+\Jdc\}\\
	\Jda &= \op{U}(s,t,0) \Big(\op{N} \{\op{U}(t',s,0) \op{L}^{(2)}_{1\,2}\} - \{\op{U}(t',s,0) \op{L}^{(2)}_{1\,2}\}\,\op{N}\Big) p_1(\vec 1,s) p_1(\vec 2,s) \\
	\Jdb &=  \op{U}(s,t,0) \sum_{i=1,2; \alpha} D_\alpha \{\partial_{i_\alpha} \op{U}(t',s,0) a_{1\,2}'\} \partial_{i_\alpha} \{\op{U}(t',s,0) \op{L}^{(2)}_{1\,2}\} p_1(\vec 1,s) p_1(\vec 2,s)\\
	\Jdc &= \op{U}(s,t,0) \{\op{N} \op{U}(t',s,0) a_{1\,2}'\} \{\op{U}(t',s,0) \op{L}^{(2)}_{1\,2}\} p_1(\vec 1,s) p_1(\vec 2,s) \text{.}
\end{align}\end{subequations}
$\Jda$ combines two terms: the noise acting on the propagated source and the noise acting on the one-particle functions alone, from the last line of \cref{eq:g2}. As both terms also lead to the same angular integrals, we treat them together throughout. Ultimately, the splitting into $\Jda$, $\Jdb$, and $\Jdc$ is artificial, and one could use the commutator form \cref{eq:Jdef} directly, particularly when using a Fourier representation.

\section{Evaluation of the Dyson-Landau collision integral} \label{app:collintegral}

In this appendix, we evaluate the three contributions $\Jda$, $\Jdb$, and $\Jdc$ to $\Jd$, \cref{eq:resultjda}, for the model \cref{eq:model} with spatially homogeneous one-particle functions. We use the angular Fourier modes $\hat p_m$ introduced in \cref{sec:daam}. Besides naturally encoding orientational order, they are the eigenmodes of the noise operator. We write $K(\theta_1,\theta_2,t)=\op{L}^{(2)}_{1\,2}\, p_1(\theta_1,t) p_1(\theta_2,t)$, whose Fourier coefficients are
\begin{align}
	\hat{K}_{mk} &= \int\frac{\opd \theta_1}{2\pi} \int\frac{\opd \theta_2}{2\pi} \exp{-\ii m \theta_1} \exp{-\ii k \theta_2}K(\theta_1,\theta_2,t) = \frac{1}{2}(m-k) \big(\hat{p}_{m+1}(t) \hat{p}_{k-1}(t)-\hat{p}_{m-1}(t) \hat{p}_{k+1}(t)\big) \text{.}
\end{align}
For homogeneous $p_1$, the free propagation does not affect the time argument of the modes. We further write $\tau_s=s-t'$ and $\tau=t-t'$. With \cref{eq:aprop}, the derivatives of the propagated indicator function with respect to the angles are
\begin{align}
\partial_{\theta_i} \op U(t',s,0)\, a_{1\,2}' &= \tau_s\, (\partial_{\theta_i} \vec{v}_{\text{rel}}) \cdot (\vec{r}_{12} - \tau_s \vec{v}_{\text{rel}})\,\delta\big(1- \lVert \vec{r}_{12}-\tau_s \vec{v}_{\text{rel}}\rVert\big) \text{,}\\
\begin{split}
\partial_{\theta_i}^2 \op U(t',s,0)\, a_{1\,2}' &= \Big\{\tau_s (\partial_{\theta_i}^2 \vec{v}_{\text{rel}}) \cdot (\vec{r}_{12}-\tau_s \vec{v}_{\text{rel}}) - \tau_s^2+\big(\tau_s (\partial_{\theta_i}\vec{v}_{\text{rel}})\cdot(\vec{r}_{12}-\tau_s \vec{v}_{\text{rel}})\big)^2\Big\}\,\delta\big(1- \lVert \vec{r}_{12}-\tau_s \vec{v}_{\text{rel}}\rVert\big) \\
&\quad+ \frac{\big\{\tau_s (\partial_{\theta_i} \vec{v}_{\text{rel}})\cdot (\vec{r}_{12}-\tau_s \vec{v}_{\text{rel}})\big\}^2}{\lVert \vec{r}_{12}-\tau_s \vec{v}_{\text{rel}}\rVert^2}\, \delta'\big(1- \lVert \vec{r}_{12}-\tau_s \vec{v}_{\text{rel}}\rVert\big) \text{.}
\end{split}
\end{align}
The subsequent propagation from $s$ to $t$ shifts $\vec{r}_{12}-\tau_s \vec{v}_{\text{rel}}$ to $\vec{r}_{12}-\tau \vec{v}_{\text{rel}}$, while the prefactors $\tau_s$ are unaffected. These prefactors then constitute the only $s$-dependence of the integrand, and $\int_{t'}^t\opd{s}\, \tau_s^n = \tau^{n+1}/(n+1)$.

\subsection{Interaction term $\Jda$}

We start with $\Jda$, which describes the smearing of the interaction kernel by the noise. Here, the indicator function is not subject to derivatives, so that the time propagation can be evaluated as in the deterministic Landau case, and the spatial integrals only yield the second moment of the collision duration, \cref{app:tsquare},
\begin{align}
\int\!\opd\vec{r}_2\, a_{1\,2} \int_{t_0}^t \opd t' \int_{t'}^t \opd s\, 1 = \int\!\opd\vec r_2\,a_{1\,2}\,\frac{(t-t_0)^2}{2} = \frac{\pi}{2v_\text{rel}^2}=\frac{\pi}{8\sin^2(\Delta/2)} \text{.}
\end{align}
In Fourier space, the noise operator is diagonal, $\op N\to m^2+k^2$, and the commutator in $\Jda$ follows directly from $\hat K_{mk}$. If we denote the Fourier transform by $\mathcal{F}$ such that $\hat{p}_k = \mathcal{F}(p_1)_k$, we find
\begin{align}
\mathcal{F}\big([\op N\bigcomma\op L^{(2)}_{1\,2}]\,p_1p_1\big)_{mk} = -(m-k)\big\{(m-k+1)\,\hat p_{m+1}\hat p_{k-1}+(m-k-1)\,\hat p_{m-1}\hat p_{k+1}\big\} \text{.}
\end{align}
The Fourier transformed contribution to the collision integral is then
\begin{align}
 \hat{\operatorname{J}}^{\text{D,1}}_\text{coll,n}&=\frac{1}{2\pi} \int_{0}^{2\pi} \exp{-\ii n \theta_1}\,  \operatorname{J}^{\text{D,1}}_\text{coll}\, \opd\theta_1 \\
 \begin{split}
 \gamma^{-2}M^{-1} \Pe\,  \hat{\operatorname{J}}^{\text{D,1}}_\text{coll,n} &= -\frac{1}{2\pi} \int_{0}^{2\pi} \opd\theta_1 \exp{-\ii n \theta_1} \partial_{\theta_1} \!\!\int\!\!\opd\theta_2\,  \sin(\theta_2-\theta_1)\, \frac{\pi}{8\sin^2(\Delta/2)} \\
 & \quad \sum_{m,k} \exp{\ii m \theta_1} \exp{\ii k \theta_2}\, \big(-(m-k)\big)\big\{(m-k+1)\,\hat{p}_{m+1}(t) \hat{p}_{k-1}(t)+(m-k-1)\,\hat{p}_{m-1}(t) \hat{p}_{k+1}(t)\big\}\end{split}\\
   \begin{split}
&=\frac{\ii n}{2\pi} \int_{0}^{2\pi} \opd\theta \!\!\int\!\!\opd\Delta\, \sum_{m,k} \exp{\ii (m+k-n) \theta} \exp{\ii k \Delta}\, \frac{\pi\sin\Delta}{8 \sin^2(\Delta/2)} \\
 &\quad (m-k)\big\{(m-k+1)\,\hat{p}_{m+1}(t) \hat{p}_{k-1}(t)+(m-k-1)\,\hat{p}_{m-1}(t) \hat{p}_{k+1}(t)\big\}\end{split}\\
   \begin{split}
 &= \ii n\sum_{k} \int\!\!\opd\Delta\, \exp{\ii k \Delta}\, \frac{\pi\sin\Delta}{8 \sin^2(\Delta/2)}\, (n-2k)\\
 &\quad\big\{(n-2k+1)\,\hat{p}_{n-k+1}(t) \hat{p}_{k-1}(t)+(n-2k-1)\,\hat{p}_{n-k-1}(t) \hat{p}_{k+1}(t)\big\}\end{split}\\
   \begin{split}
 &= \ii n \sum_{k}  \hat{p}_{n-k}(t) \hat{p}_{k}(t)\int\!\!\opd\Delta\, \exp{\ii k \Delta}\, \frac{\pi\sin\Delta}{8 \sin^2(\Delta/2)} \\
 &\quad \big\{(n-2k-2)(n-2k-1) \exp{\ii \Delta} + (n-2k+2)(n-2k+1) \exp{-\ii \Delta} \big\}\end{split} \\
 &= \ii n \sum_{k}  \hat{p}_{n-k}(t) \hat{p}_{k}(t)\int\!\!\opd\Delta\, \exp{\ii k \Delta}\, \frac{\pi\sin\Delta}{8 \sin^2(\Delta/2)}\, \Big\{ 2\big((n-2k)^2+2\big) \cos\Delta - 6\ii (n-2k) \sin\Delta \Big\} \text{.}
\end{align}
This leaves two integrals to be evaluated, namely
\begin{align}
I_1&=\int\!\!\opd\Delta\, \exp{\ii k \Delta}\,  \frac{\sin\Delta\cos\Delta}{\sin^2(\Delta/2)}=2\pi \ii \sgn(k) (2-\delta_{k,1}-\delta_{k,-1}) \text{,}\\
I_2&=\int\!\!\opd\Delta\, \exp{\ii k \Delta}\, \frac{\sin^2\Delta}{\sin^2(\Delta/2)} = 4 \int\!\!\opd\Delta\, \exp{\ii k \Delta} \cos^2(\Delta/2)= 2 \int\!\!\opd\Delta\, \exp{\ii k \Delta} (1+\cos\Delta)=2\pi(2\delta_{k,0}+\delta_{k,1}+\delta_{k,-1}) \text{.}
\end{align}
The integrand of $I_1$ has an odd singularity at $\Delta=0$, so that its principal value exists. We show details in \cref{app:deriv}. Using these integrals, we find
\begin{align}
	\begin{split}
 \gamma^{-2}M^{-1} \Pe\,  \hat{\operatorname{J}}^{\text{D,1}}_\text{coll,n} &= \frac{n\pi^2}{2} \sum_{k}  \hat{p}_{n-k}(t) \hat{p}_{k}(t) \\&\quad \Big\{ 6n\,\delta_{0,k} + \big(((n-2k)^2+2)\sgn(k) + 3(n-2k)\big)\delta_{1,\lvert k\rvert} {-} 2\big((n-2k)^2+2\big)\sgn(k)\Big\} \text{.} \end{split}
\end{align}

\subsection{Smearing term $\Jdc$}

For the geometry smearing term,
\begin{align}
\Jdc &=   \{\op{U}(s,t,0)\op N \op{U}(t',s,0) a_{1\,2}'\}\, K(\theta_1,\theta_2,t) \equiv \tildeJdc\, K(\theta_1,\theta_2,t) \text{,}
\end{align}
we find with $\Delta = \theta_2-\theta_1$ and $\sum_{i}\partial^2_{\theta_i}\vec v_\text{rel}=-\vec v_\text{rel}$
\begin{align}
\begin{split}
\int_{t'}^t\opd{s}\,\tildeJdc ={}& -\Big\{-\frac{\tau^2}{2}\, \vec{v}_{\text{rel}} \cdot (\vec{r}_{12}-\tau \vec{v}_{\text{rel}}) - \frac{2\tau^3}{3}+\frac{\tau^3}{3} \sum_{i=1,2} \big( (\partial_{\theta_i}\vec{v}_{\text{rel}})\cdot(\vec{r}_{12}-\tau \vec{v}_{\text{rel}})\big)^2\Big\}\,\delta\big(1- \lVert \vec{r}_{12}-\tau \vec{v}_{\text{rel}}\rVert\big) \\
&- \sum_{i=1,2}\frac{\tau^3}{3}\,\frac{\big\{(\partial_{\theta_i} \vec{v}_{\text{rel}})\cdot (\vec{r}_{12}-\tau \vec{v}_{\text{rel}})\big\}^2}{\lVert \vec{r}_{12}-\tau \vec{v}_{\text{rel}}\rVert^2}\, \delta'\big(1- \lVert \vec{r}_{12}-\tau \vec{v}_{\text{rel}}\rVert\big) \text{.}
\end{split}
\end{align}
We can write the collision condition as $r_{12}= \tau v_{\text{rel}}\cos\phi + \sqrt{v_{\text{rel}}^2\cos^2\phi\,\tau^2+1-\tau^2v_{\text{rel}}^2} \equiv r_c$, wherein we use $\cos\phi=(\vec{v}_{\text{rel}}\cdot \vec{r}_{12})/(v_{\text{rel}} r_{12})$. This is beneficial as it moves the poles of the delta functions to the interior of the $r_{12}$ integration. We know that $0\leq r_{12}\leq 1$, which leads to $0\leq \tau \leq 2 \cos\phi/v_{\text{rel}}$. Thus, there are only contributions in the receding case, $\cos\phi>0$. We state familiar properties of Dirac's delta distribution with some function $f(x)$ that has a single zero at $x_0$,
\begin{align}
\int\opd{x}\, g(x) \delta(f(x)) &= g(x_0)/\lvert f'(x_0)\rvert \text{,}\\
\int\opd{x}\, g(x) \delta'(f(x)) &= - \frac{1}{\lvert f'(x_0)\rvert f'(x_0)} \Big( g'(x_0) + \frac{g(x_0) f''(x_0)}{f'(x_0)}\Big) \text{.}
\end{align}
Naturally, these only hold if $x_0$ is within the integration domain. To avoid having to evaluate the curvature contributions in the second line, we shift the $\vec{r}_2$ integration, $\vec{r}_2 \to \vec{q}=\vec{r}_{12}-\tau\vec{v}_{\text{rel}}$. We write $\vec{q}=q(\cos{\alpha},\sin\alpha)$ in a frame in which $-\vec v_\text{rel}$ points along the first axis, which leads to $\vec{v}_{\text{rel}}\cdot \vec q=-v_{\text{rel}}\, q  \cos\alpha$. As we define $\alpha$ relative to the relative velocity, there is a hidden dependence on the angles $\theta_i$ that has to be considered for the terms with derivatives. In a fixed frame of reference, we write $\vec{q}=q (\cos\beta,\sin\beta)$ and, with $c=\theta_1+\theta_2$, find $\vec{v}_{\text{rel}}\cdot \vec q = 2 \sin(\Delta/2)\, q \sin(\beta-c/2)$ as well as $(\partial_{\theta_i}\vec{v}_{\text{rel}})\cdot \vec q = (\partial_{\theta_i}\Delta ) \cos(\Delta/2)\, q \sin(\beta-c/2)-\sin(\Delta/2)\,q \cos(\beta-c/2)$. For the relevant sum of squares, this means $\sum_i ((\partial_{\theta_i} \vec{v}_{\text{rel}}) \cdot \vec{q})^2=q^2(1-\cos\Delta\cos(2\beta-c))=q^2 (1+\cos\Delta\cos(2\alpha))$, and thus
\begin{align}
\begin{split}
\int_{t'}^t\opd{s}\,\tildeJdc={}& \Big\{-\frac{\tau^2}{2} v_{\text{rel}}\, q \cos\alpha+ \frac{2 \tau^3}{3} - \frac{\tau^3 q^2}{3} \big(1+\cos\Delta\cos(2\alpha)\big)\Big\} \delta(1- q) \\
&- \frac{\tau^3}{3} \big(1+\cos\Delta\cos(2\alpha)\big) \delta'(1- q) \text{,}
\end{split}\\
\int\opd{\vec{r}_2}\, a_{12} \int_{t'}^t\opd{s}\,\tildeJdc &=  \int\opd{\alpha}\, \Theta(\cos\phi)\, \Theta(2\cos\phi/v_{\text{rel}}-\tau) \int\opd{q}\, q  \int_{t'}^t\opd{s}\,\tildeJdc \text{.}
\end{align}
We can rewrite the $\Theta$-constraints as
\begin{align}
\Theta(\cos\phi)\,\Theta(2\cos\phi/v_{\text{rel}}-\tau)\,\Theta(r_{12}) \to \Theta(\cos\alpha)\, \Theta(\cos\alpha-\tau v_{\text{rel}}/2) \text{,}
\end{align}
which confirms that they are indeed independent of $q$. More importantly, it shows that the spatial dependence in the eventual $\tau$-integration is such that it was justified to exchange the order of integrations. Performing the $q$-integration gives
\begin{align}
\int\opd{q}\, q \int_{t'}^t\opd{s}\,\tildeJdc &= -  \frac{\tau^2}{2} v_{\text{rel}}  \cos\alpha - \frac{2 \tau^3}{3} \cos\Delta\cos(2\alpha) \text{.}
\end{align}
In this form, we can now safely perform the integral $\int_{t_0}^t \opd{t'}=\int_0^{t-t_0} \opd{\tau} \to \int_0^{2\cos\alpha/v_{\text{rel}}} \opd{\tau}$,
\begin{align}
\int_0^{2\cos\alpha/v_{\text{rel}}} \!\!\!\!\!\!\!\!\!\opd{\tau}\int\opd{q}\, q \int_{t'}^t\opd{s}\,\tildeJdc &= \Big(-  \frac{\tau^3}{6} v_{\text{rel}}  \cos\alpha - \frac{\tau^4}{6} \cos\Delta\, ( 2\cos^2\alpha-1)\Big)\Big\rvert_{\tau=2\cos\alpha/v_\text{rel}} \text{.}
\end{align}
We can directly perform the $\alpha$-integration as well using $\int_{-\pi/2}^{\pi/2}\opd x\, \cos^n x= \frac{\sqrt{\pi}\,\Gamma((n+1)/2)}{\Gamma(1+n/2)}=B((n+1)/2,1/2)$ with the Gamma and Beta functions, respectively. Combining everything, we get
\begin{align}
\int_{-\pi}^\pi \opd{\alpha}\, \Theta(\cos\alpha)\int_0^{2\cos\alpha/v_{\text{rel}}} \!\!\!\!\!\!\!\!\!\opd{\tau}\int\opd{q}\, q \int_{t'}^t\opd{s}\,\tildeJdc &= -  \frac{\pi}{24} \frac{1+\sin^2(\Delta/2)}{\sin^4(\Delta/2)} \text{.}
\end{align}

We write
\begin{align}
\operatorname{J}^{\text{D,3}}_\text{coll}  &= -\gamma^2 M\Pe^{-1} \,\partial_{\theta_1} \!\!\int\!\!\opd \vec{2}\, a_{1\,2} F_{1\,2}  \int_{t_0}^t \opd t'  \int_{t'}^t\opd{s}\, \Jdc \text{,}\\
 \hat{\operatorname{J}}^{\text{D,3}}_\text{coll,n}&=\frac{1}{2\pi} \int_{0}^{2\pi} \exp{-\ii n \theta_1}\,  \operatorname{J}^{\text{D,3}}_\text{coll}\, \opd\theta_1 \text{,}
\end{align}
and can apply our findings to give
\begin{align}
 \begin{split}
 \gamma^{-2}M^{-1} \Pe\,  \hat{\operatorname{J}}^{\text{D,3}}_\text{coll,n} &= \frac{1}{2} \frac{1}{2\pi} \int_{0}^{2\pi} \opd\theta_1 \exp{-\ii n \theta_1} \partial_{\theta_1} \!\!\int\!\!\opd\theta_2\,  \sin\Delta\,  \frac{\pi}{24} \frac{1+\sin^2(\Delta/2)}{\sin^4(\Delta/2)} \\
 & \quad \sum_{m,k} \exp{\ii m \theta_1} \exp{\ii k \theta_2} (m-k) \big(\hat{p}_{m+1}(t) \hat{p}_{k-1}(t)-\hat{p}_{m-1}(t) \hat{p}_{k+1}(t)\big)\end{split}\\
  \begin{split}
 &= \frac{1}{2} \frac{1}{2\pi} \int_{0}^{2\pi} \opd\theta \int\!\!\opd\Delta\,\sum_{m,k} \exp{\ii (m+k-n) \theta} \exp{\ii k \Delta}\,  \sin\Delta\,  \frac{\pi}{24} \frac{1+\sin^2(\Delta/2)}{\sin^4(\Delta/2)} \\
 & \quad \ii n(m-k) \big(\hat{p}_{m+1}(t) \hat{p}_{k-1}(t)-\hat{p}_{m-1}(t) \hat{p}_{k+1}(t)\big)\end{split}\\
  \begin{split}
 &= \frac{1}{2} \int\!\!\opd\Delta\,\sum_{k}  \exp{\ii k \Delta}\,  \sin\Delta\,  \frac{\pi}{24} \frac{1+\sin^2(\Delta/2)}{\sin^4(\Delta/2)} \\
 & \quad \ii n(n-2k) \big(\hat{p}_{n-k+1}(t) \hat{p}_{k-1}(t)-\hat{p}_{n-k-1}(t) \hat{p}_{k+1}(t)\big)\end{split}
 \\
  \begin{split}
 &= \frac{\ii n}{2}  \sum_{k} \hat p_{n-k}(t) \hat p_k(t) \int\!\!\opd\Delta\,  \exp{\ii k \Delta}\,  \sin\Delta\,  \frac{\pi}{24} \frac{1+\sin^2(\Delta/2)}{\sin^4(\Delta/2)} \\
 & \quad  \big((n-2k-2) \exp{\ii\Delta} -(n-2k+2)\exp{-\ii\Delta}\big)\end{split} \\
  \begin{split}
 &= \ii n \sum_{k} \hat p_{n-k}(t) \hat p_k(t) \int\!\!\opd\Delta\,  \exp{\ii k \Delta}\,  \sin\Delta\,  \frac{\pi}{24} \frac{1+\sin^2(\Delta/2)}{\sin^4(\Delta/2)} \\
 & \quad  \big(-2 \cos\Delta + \ii (n-2k) \sin{\Delta}\big) \text{.}\end{split}
\end{align}
Using the trigonometric relations $\cos\Delta\sin\Delta\,(1+\sin^2(\Delta/2))=\frac{3}{2}\sin\Delta\cos\Delta-\frac{1}{2} \cos^2\Delta\sin\Delta$ and $\sin^2\Delta\,(1+\sin^2(\Delta/2))=\frac{3}{2}\sin^2\Delta-\frac{1}{2}\cos\Delta\sin^2\Delta$, there are thus four new integrals to be evaluated,
\begin{align}
I_3&=\int\!\!\opd\Delta\, \exp{\ii k \Delta}\, \frac{\sin\Delta\cos\Delta}{\sin^4(\Delta/2)}= -8\pi (1+k^2)\, \ii \sgn(k) \text{,}\\
I_4&=\int\!\!\opd\Delta\, \exp{\ii k \Delta}\, \frac{\cos^2\Delta\sin\Delta}{\sin^4(\Delta/2)}=-8 \pi \ii (2+k^2) \sgn(k) + 4\pi \ii \sgn(k)\, \delta_{1,\lvert k \rvert} \text{,}\\
I_5&=\int\!\!\opd\Delta\, \exp{\ii k \Delta}\, \frac{\sin^2\Delta}{\sin^4(\Delta/2)}=- 16 \pi k \sgn(k) - 8\pi \delta_{0,k} \text{,}\\
I_6&=\int\!\!\opd\Delta\, \exp{\ii k \Delta}\, \frac{\cos\Delta\sin^2\Delta}{\sin^4(\Delta/2)}= - 16 \pi k \sgn(k) - 16\pi \delta_{0,k} -4\pi \delta_{1,\lvert k \rvert} \text{.}
\end{align}
These integrals do not converge at $\Delta=0$. The given values are their finite parts, obtained by contour integration as shown in \cref{app:deriv}. In contrast to the mixed term below, the divergent parts do not cancel in the sum over $k$. As discussed in \cref{sec:freenoise}, this reflects the breakdown of the first-order Dyson expansion for nearly parallel pairs. For the finite part of the collision integral, this yields
\begin{align}
\gamma^{-2}M^{-1} \Pe\,  \hat{\operatorname{J}}^{\text{D,3}}_\text{coll,n} &=  \ii n \frac{\pi}{24} \sum_{k} \hat p_{n-k}(t) \hat p_k(t) \Big(-3 I_3 + I_4 + \frac{\ii (n-2k)}{2} (3 I_5-I_6) \Big) \\
&= - \frac{n \pi^2}{12}\sum_{k} \hat p_{n-k}(t) \hat p_k(t)  \Big(-2n\, \delta_{0,k} + \big(n-2k+2\sgn(k)\big)\delta_{1,\lvert k\rvert}+4(6k^2-2kn+1)\sgn(k)\Big) \text{.}
\end{align}
The coupling coefficients are real, and the linear terms in front of $\hat p_0\hat p_n$ are damping.

\subsection{Mixed term $\Jdb$}

The mixed term distributes the second derivatives of the noise operator. For the model \cref{eq:model}, it reads
\begin{align}
\Jdb &=  \op{U}(s,t,0) \sum_{i=1,2} \{\partial_{\theta_i} \op{U}(t',s,0) a_{1\,2}'\}\, \partial_{\theta_i} K(\theta_1,\theta_2,s) \text{.}
\end{align}
With the first derivative of the propagated indicator function from above, we find
\begin{align}
 \op{U}(s,t,0)\,\partial_{\theta_i} \op U(t',s,0)\, a_{1\,2}' &= \tau_s\, (\partial_{\theta_i} \vec{v}_{\text{rel}}) \cdot (\vec{r}_{12} - \tau \vec{v}_{\text{rel}})\,\delta\big(1- \lVert \vec{r}_{12}-\tau \vec{v}_{\text{rel}}\rVert\big) \text{,}\\
\int_{t'}^t \opd{s}\, \op{U}(s,t,0)\,\partial_{\theta_i} \op U(t',s,0)\, a_{1\,2}' &= \frac{\tau^2 q}{2}\big\{(\partial_{\theta_i}\Delta ) \cos(\Delta/2)  \sin(\beta-c/2)-\sin(\Delta/2)\cos(\beta-c/2)\big\}\, \delta(1- q) \text{.}
\end{align}
With $\alpha$ measured from $-\vec v_\text{rel}$, we have $\sin(\beta-c/2)=-\sgn(\sin(\Delta/2)) \cos\alpha$, whereas the term with $\cos(\beta-c/2)$ is odd in $\alpha$ and vanishes upon integration over the symmetric domain. With $v_\text{rel}=2\lvert\sin(\Delta/2)\rvert$, this leads to
\begin{align}
\int_{-\pi}^\pi \opd{\alpha}\, \Theta(\cos\alpha)\int_0^{2\cos\alpha/v_{\text{rel}}} \!\!\!\!\!\!\!\!\!\opd{\tau}\int\opd{q}\,q \int_{t'}^t \opd{s}\, \op{U}(s,t,0)\,\partial_{\theta_i} \op U(t',s,0)\, a_{1\,2}'  &= -\frac{\pi}{16}\, (\partial_{\theta_i}\Delta )\,\frac{\cos(\Delta/2)}{\sin^3(\Delta/2)} \text{.}
\end{align}
This can be checked independently: up to the factor $\tau_s/\tau$, whose $s$-integral gives the weight $\tau/2$, the integrand is the angular derivative of the propagated indicator function, so that the result equals $\partial_{\theta_i}$ of $\int\opd\vec r_2\,a_{1\,2}\int\opd t'\,(t-t')/2=\pi/(4v_\text{rel}^2)=\pi/(16\sin^2(\Delta/2))$, cf.\ \cref{app:tsquare}. With $\partial_{\theta_1}\Delta=-1$ and $\partial_{\theta_2}\Delta=1$, we find
\begin{align}
\int\!\!\opd \vec{r}_2\, a_{1\,2} \int_{t_0}^t \opd t'  \int_{t'}^t\opd{s}\, \Jdb  &= \frac{\pi}{16}\,\frac{\cos(\Delta/2)}{\sin^3(\Delta/2)}\, (\partial_{\theta_1}-\partial_{\theta_2}) K(\theta_1,\theta_2,t)
\end{align}
and can proceed to compute the corresponding contribution to the collision integral. According to \cref{eq:resultjda}, it enters with the factor $-2$,
\begin{align}
\operatorname{J}^{\text{D,2}}_\text{coll}  &= 2\gamma^2 M\Pe^{-1} \,\partial_{\theta_1} \!\!\int\!\!\opd \vec{2}\, a_{1\,2} F_{1\,2}  \int_{t_0}^t \opd t'  \int_{t'}^t\opd{s}\, \Jdb \text{,}\\
 \hat{\operatorname{J}}^{\text{D,2}}_\text{coll,n}&=\frac{1}{2\pi} \int_{0}^{2\pi} \exp{-\ii n \theta_1}\,  \operatorname{J}^{\text{D,2}}_\text{coll}\, \opd\theta_1 \text{,}\\
 \begin{split}
 \gamma^{-2}M^{-1}\Pe\,\hat{\operatorname{J}}^{\text{D,2}}_\text{coll,n}&=-\frac{1}{32} \int_0^{2\pi} \opd{\theta} \int_{0}^{2\pi}\opd{\Delta} \sum_{m,k} \exp{\ii (m+k-n)\theta} \exp{\ii k \Delta}\, \frac{\sin\Delta\cos(\Delta/2)}{\sin^3(\Delta/2)}\\
 &\quad n(m-k)^2 \big(\hat{p}_{m+1}(t) \hat{p}_{k-1}(t)-\hat{p}_{m-1}(t) \hat{p}_{k+1}(t)\big) \end{split}
 \\
 \begin{split}
&=-\frac{\pi n}{16}\int_{0}^{2\pi}\opd{\Delta} \sum_{k} \exp{\ii k \Delta}\,\frac{\sin\Delta\cos(\Delta/2)}{\sin^3(\Delta/2)}\, (n-2k)^2 \big(\hat{p}_{n-k+1}(t) \hat{p}_{k-1}(t)-\hat{p}_{n-k-1}(t) \hat{p}_{k+1}(t)\big) \end{split} \\
 \begin{split}
&=-\frac{\pi n}{16}\sum_{k} \hat{p}_{n-k}(t) \hat{p}_{k}(t) \int_{0}^{2\pi}\opd{\Delta}\,  \exp{\ii k \Delta}\,\frac{\sin\Delta\cos(\Delta/2)}{\sin^3(\Delta/2)}\, \big((n-2k-2)^2 \exp{\ii\Delta}-(n-2k+2)^2\exp{-\ii\Delta} \big) \end{split}\\
&=-\frac{\pi n}{16}\sum_{k} \hat{p}_{n-k}(t) \hat{p}_{k}(t) \int_{0}^{2\pi}\opd{\Delta}\,  \exp{\ii k \Delta}\,\frac{\sin\Delta\cos(\Delta/2)}{\sin^3(\Delta/2)}\, \big(-8(n-2k)\cos\Delta+2\ii((n-2k)^2+4)\sin\Delta \big) \text{.}
\end{align}
Thus, two new integrals are to be evaluated,
\begin{align}
I_7&=\int\!\!\opd\Delta\, \exp{\ii k \Delta}\, \frac{\sin\Delta\cos\Delta\cos(\Delta/2)}{\sin^3(\Delta/2)}= - 8 \pi k \sgn(k) - 8\pi \delta_{0,k} -2\pi \delta_{1,\lvert k \rvert} \text{,}\\
I_8&=\int\!\!\opd\Delta\, \exp{\ii k \Delta}\, \frac{\sin^2\Delta\cos(\Delta/2)}{\sin^3(\Delta/2)} = 8\pi \ii \sgn(k) - 2\pi \ii \sgn(k)\, \delta_{1,\lvert k \rvert} \text{,}
\end{align}
for which we provide details in \cref{app:deriv}. The integrand of $I_8$ has an odd singularity, so that its principal value exists. $I_7$ exists only as a finite part and its divergent part is independent of $k$ and enters with the factor $n-2k$, so that it cancels in the sum over $k$, as $\hat p_{n-k}\hat p_k$ is symmetric under $k\to n-k$. For the contribution to the collision integral, we get
\begin{align}
\gamma^{-2}M^{-1}\Pe\,\hat{\operatorname{J}}^{\text{D,2}}_\text{coll,n}&= -\frac{\pi n }{16} \sum_{k} \hat{p}_{n-k}(t) \hat{p}_{k}(t) \big(-8(n-2k) I_7 + 2 \ii ((n-2k)^2+4)I_8\big)\\
\begin{split}
&=-\frac{\pi^2 n}{4}\sum_{k}\hat{p}_{n-k}(t)\,\hat{p}_{k}(t)\\
&\quad\Big\{16n\,\delta_{0,k}
+\big(4 (n-2k) + ((n-2k)^2+4)\sgn(k)\big)\delta_{1,\lvert k\rvert}
-4 \big((n-2k)(n-6k)+4\big)\sgn(k)\Big\} \text{.} \end{split}
\end{align}
The coupling coefficients are real. In contrast to the other two contributions, the linear terms in front of $\hat{p}_{0} \hat{p}_{n}$ are anti-damping.

\section{Computation of $\tau^2$} \label{app:tsquare}
In this section, we compute the second moment of the time elapsed since the beginning of the collision, averaged over the interaction region, for free active particles,
\begin{align}
\begin{split}
\tau^2&=\frac{1}{\pi} \int_{-1}^1 \opd y \int_{-\sqrt{1-y^2}}^{\sqrt{1-y^2}}\!\!\opd x \left(\frac{x+\sqrt{1-y^2}}{v_\text{rel}} \right)^2 =\frac{2}{\pi v_{\text{rel}}^2} \int_0^1 \frac{8}{3}\,(1-y^2)^{3/2}\, \opd y=\frac{16}{3\pi v_{\text{rel}}^2} \int_0^{\frac{\pi}{2}} \cos^4 z\, \opd z\\
&=\frac{2}{3\pi v_{\text{rel}}^2} \int_0^{\frac{\pi}{2}} \big(\cos(4z)+4\cos(2z)+3\big)\, \opd z=\frac{2}{3\pi v_{\text{rel}}^2}\,\frac{3\pi}{2}=\frac{1}{v_{\text{rel}}^2} \text{,}
\end{split}
\end{align}
where we substituted $y=\sin z$. For details on this computation, see the discussion of the first moment in Refs.~\cite{ihle2023,boltz2024}.

\section{Noise integrals for gated noise} \label{app:gated}

For homogeneous one-particle functions, both noise integrals in \cref{eq:modesnoise} only involve the deterministic collision duration integrated over the interaction region, $\int\opd\vec r_2\,a_{1\,2}\int_{t_0}^t\opd t'\,1 = 8/(3v_\text{rel})=4/(3\lvert\sin(\Delta/2)\rvert)$~\cite{boltz2024}, and the Fourier coefficients $\hat K_{mk}$ from \cref{app:collintegral}.

\subsection{Integral $X$}

With $\big((\op N_1+\op N_2)\,p_1p_1\big)$ corresponding to $(m^2+k^2)\,\hat p_m\hat p_k$ in Fourier space, we find
\begin{align}
\begin{split}
\hat X_n &= \frac{1}{2\pi}\int\opd\theta_1\,\exp{-\ii n\theta_1}\,\partial_{\theta_1}\!\int\!\opd\theta_2\,\sin\Delta\,\frac{4}{3\lvert\sin(\Delta/2)\rvert}\sum_{m,k}\exp{\ii m\theta_1}\exp{\ii k\theta_2}\,(m^2+k^2)\,\hat p_m(t)\hat p_k(t)
\end{split}\\
&= \frac{4\ii n}{3}\sum_k \big((n-k)^2+k^2\big)\,\hat p_{n-k}(t)\hat p_k(t)\, I_{10}(k) \text{,}
\end{align}
with
\begin{align}
I_{10}(k)=\int\opd\Delta\,\exp{\ii k\Delta}\,\frac{\sin\Delta}{\lvert\sin(\Delta/2)\rvert} = 4\ii\int_0^\pi\opd\Delta\,\sin(k\Delta)\cos(\Delta/2)=\frac{16\ii k}{4k^2-1} \text{,}
\end{align}
where we used $\sin\Delta/\lvert\sin(\Delta/2)\rvert=2\sgn(\Delta)\cos(\Delta/2)$ on $(-\pi,\pi)$. Symmetrizing the coefficients under $k\to n-k$ gives \cref{eq:Xhat}. The integrand is bounded, so that a regularization of the collision duration only affects $\hat X_n$ at order $\mathcal O(\delta)$.

\subsection{Integral $Y$}

Using $\hat K_{mk}$, we find
\begin{align}
\hat{Y}_n &= - \frac{1}{3 \pi} \int\opd{\theta_1}\, \exp{-\ii n \theta_1}\, \partial_{\theta_1}^2 \!\int\!\opd{\theta_2} \sum_{m,k} \frac{\exp{\ii m \theta_1} \exp{\ii k \theta_2}}{\lvert \sin(\Delta/2)\rvert} (m-k) \big(\hat{p}_{m+1}(t) \hat{p}_{k-1}(t)-\hat{p}_{m-1}(t) \hat{p}_{k+1}(t)\big)\\
&=  \frac{n^2}{3 \pi} \int\opd{\theta} \sum_{m,k} \int\opd{\Delta}\, \exp{\ii (m+k-n) \theta} \exp{\ii k \Delta}\, \frac{m-k}{\lvert \sin(\Delta/2)\rvert} \big(\hat{p}_{m+1}(t) \hat{p}_{k-1}(t)-\hat{p}_{m-1}(t) \hat{p}_{k+1}(t)\big) \\
&=  \frac{2 n^2}{3} \sum_{k} \int\opd{\Delta}\, \exp{\ii k \Delta}\, \frac{n-2k}{\lvert \sin(\Delta/2)\rvert} \big(\hat{p}_{n-k+1}(t) \hat{p}_{k-1}(t)-\hat{p}_{n-k-1}(t) \hat{p}_{k+1}(t)\big)\\
&= \frac{2 n^2}{3} \sum_{k} \hat{p}_{n-k}(t) \hat{p}_k(t) \int\opd{\Delta}\, \frac{\exp{\ii k \Delta}}{\lvert \sin(\Delta/2)\rvert} \big(-4 \cos \Delta + 2\ii (n-2k)\sin\Delta\big) \text{,}
\end{align}
which with $I_9$ and $I_{10}$ gives \cref{eq:Yhat}. The integrand of $I_9$ is even, so that its imaginary part vanishes, whereas its real part diverges logarithmically at $\Delta=0$.

\section{Evaluation of the angular integrals} \label{app:deriv}

\begin{figure}
\includegraphics{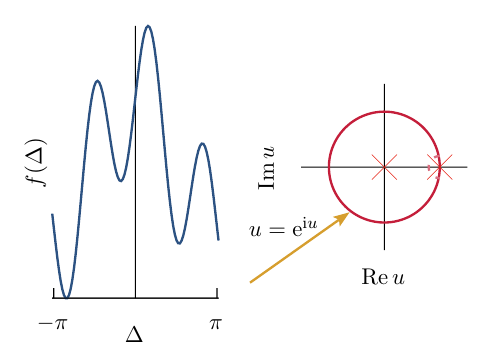}
\caption{Sketch of the integration method. An integral over a periodic function is transformed into a contour integral along the unit circle in the complex plane and evaluated by means of the residue theorem. A pole on the contour is excluded by a small indentation and contributes half of its residue. This yields the Cauchy principal value or, for poles of higher order, the Hadamard finite part of integrals that do not exist as improper Riemann integrals.}
\label{fig:sketch}
\end{figure}

\subsection{Method}

All angular integrals $I_1$ to $I_8$ are of the form $\int_{-\pi}^{\pi}\opd\Delta\,\exp{\ii l\Delta}F(\Delta)$, where $F$ is $2\pi$-periodic, rational in $\exp{\ii\Delta}$ after simplification, and has a pole at $\Delta=0$ only. With $u=\exp{\ii\Delta}$ and $\opd\Delta=\opd u/(\ii u)$, they become contour integrals $\oint_{\abs{u}=1}f(u)\,\opd u$ of rational functions $f$, which have poles at $u=0$, inside the contour, and at $u=1$, on the contour, see \cref{fig:sketch}.

If $F$ has a simple pole at $\Delta=0$, as for $I_1$ and $I_8$, the integral does not exist as an improper Riemann integral, but its Cauchy principal value does. Excluding $u=1$ by a small indentation of the contour, the pole contributes half of its residue~\cite{ahlfors1979},
\begin{align}
\int_{-\pi}^{\pi}\opd\Delta\,\exp{\ii l\Delta}F(\Delta) = \pi\ii\Res(f,1) + 2\pi\ii\Res(f,0) \text{.} \label{eq:halfres}
\end{align}
The principal value is the limit of a symmetrically regularized integral, e.g.\ of a slight difference of the speeds of the two particles, which renders $v_\text{rel}$ finite for $\Delta=0$.

If $F$ has a pole of higher order, as for $I_3$ to $I_7$, the even parts of its Laurent expansion, $F(\Delta)=\sum_{j\geq1} b_j\Delta^{-j}+\text{regular}$, render the integral divergent, and only its Hadamard finite part is defined~\cite{hadamard1923,gelfand1964},
\begin{align}
\operatorname{FP}\!\int_{-\pi}^{\pi}\opd\Delta\,F(\Delta) = \lim_{\eps\to0}\bigg\{\int_{\eps<\abs{\Delta}<\pi}\opd\Delta\,F(\Delta) - \sum_{j\ \text{even}} \frac{2b_j}{j-1}\,\eps^{1-j}\bigg\} \text{.}
\end{align}
\Cref{eq:halfres} also holds for the finite part. On the unit circle, the excised integral of $(u-1)^{-n}$ with $n\geq2$ is an odd function of $\eps$ and thus has no $\eps$-independent part, whereas $n=1$ yields $\pi\ii$. The regular part of $f$ at $u=1$ contributes its residues inside the contour. Wherever the principal value exists, it coincides with the finite part. For the regularization $\sin^2(\Delta/2)\to\sin^2(\Delta/2)+\delta^2$ of the singular denominators, the regularized integrals behave as $c/\delta+\operatorname{FP}+\mathcal O(\delta)$. In \cref{sec:freenoise}, the divergent parts of $I_3$ to $I_6$ give rise to the non-analytic contribution, and the finite parts to the remaining regular contribution. The divergent part of $I_7$ cancels in the sum over the Fourier index, see \cref{app:collintegral}.

In the following, $\Theta(0)=0$ and $\sgn(0)=0$.

\subsection{Integrals with simple poles}

We first discuss $I_1$ in detail. With
\begin{align}
\sin\Delta \cos\Delta = \frac{u^4-1}{4\ii u^2} \text{,}\qquad
\sin^2\left(\frac{\Delta}{2}\right) = - \frac{(u-1)^2}{4 u} \text{,}
\end{align}
we find
\begin{align}
I_1 = \int_{-\pi}^{\pi}\opd{\Delta}\, \exp{\ii l \Delta}\,\frac{\sin\Delta \cos\Delta}{\sin^2(\Delta/2)} = \oint_{\abs{u}=1} f_1(u)\,\opd{u} \text{,}\qquad f_1(u) = u^{l-2}\, \frac{(u^2+1)(u+1)}{u-1} \text{.}
\end{align}
The integrand has a simple pole at $u=1$ with $\Res(f_1,1)=4$. For $l<2$, it additionally has a pole at $u=0$. Its residue follows from the Laurent series,
\begin{align}
f_1(u) = - \big(u^{l+1} + u^l + u^{l-1} + u^{l-2}\big) \sum_{n=0}^\infty u^n \text{,}
\end{align}
in which each of the four terms contributes $-1$ to the coefficient of $u^{-1}$ if $n=-l-2$, $-l-1$, $-l$, or $1-l$, respectively, is non-negative. Hence, $\Res(f_1,0)=-\max\big(0,\min(4,2-l)\big)$, and \cref{eq:halfres} gives
\begin{align}
I_1 = 4\pi\ii + 2\pi\ii\Res(f_1,0) = 2\pi\ii\times\begin{cases} 2 & l\geq 2\\ 1 & l=1\\ 0 & l=0\\ -1 & l=-1\\ -2 & l\leq -2\end{cases} \quad = 2\pi \ii \sgn(l) \big(2 - \delta_{\abs{l}, 1}\big) \text{.}
\end{align}

For $I_8$, the integrand simplifies with $\sin\Delta=2\sin(\Delta/2)\cos(\Delta/2)$ to $4\cos^3(\Delta/2)/\sin(\Delta/2)$, which is odd in $\Delta$. Combining $\Delta$ and $-\Delta$, the singularity cancels,
\begin{align}
I_8=\int\!\!\opd\Delta\, \exp{\ii l \Delta}\, \frac{\sin^2\Delta\cos(\Delta/2)}{\sin^3(\Delta/2)}= 8 \ii \int_0^{\pi}\!\!\opd\Delta\, \sin(l \Delta)\, \frac{\cos^3(\Delta/2)}{\sin(\Delta/2)}
= 8\pi \ii \sgn(l) - 2\pi \ii \sgn(l)\, \delta_{1,\abs{l}} \text{,}
\end{align}
where we used $\cos^3x/\sin x=\cot x-\frac12\sin(2x)$ and $\int_0^\pi\opd\Delta\,\sin(l\Delta)\cot(\Delta/2)=\pi$ for $l\geq1$. \Cref{eq:halfres} gives the same result.

\subsection{Integrals with poles of higher order}

The remaining integrals are evaluated in the same way, we only give the integrands and residues. For $I_3$, we find
\begin{align}
I_3 &= \int\!\!\opd\Delta\, \exp{\ii l \Delta}\, \frac{\sin\Delta\cos\Delta}{\sin^4(\Delta/2)}= \oint_{\abs{u}=1} f_3(u)\, \opd{u} \text{,}\qquad f_3(u)=\frac{4u^{l-1}(1+u)(1+u^2)}{(1-u)^3} \text{,}
\end{align}
with $\Res(f_3,1)=-8(1+l^2)$ and $\Res(f_3,0)=8(1+l^2)\Theta(-l)+4\delta_{0,l}$. Combining these gives
\begin{align}
I_3 &= \pi \ii \big(\Res(f_3,1)+2 \Res(f_3,0)\big) = -8\pi (1+l^2)\, \ii \sgn(l) \text{.}
\end{align}
For $I_4$, we find
\begin{align}
I_4 &= \int\!\!\opd\Delta\, \exp{\ii l \Delta}\, \frac{\sin\Delta\cos^2\Delta}{\sin^4(\Delta/2)}=  \oint_{\abs{u}=1} f_4(u)\,  \opd{u} \text{,}\qquad f_4(u)=\frac{2 (u+1) \left(u^2+1\right)^2 u^{l-2}}{(1-u)^3} \text{,}
\end{align}
with $\Res(f_4,1)=-8(2+l^2)$ and $\Res(f_4,0)=8(2+l^2)\Theta(-l)+2 \sgn(l)\,\delta_{1,\abs{l}}+8\delta_{0,l}$, which gives
\begin{align}
I_4 &= -8 \pi \ii (2+l^2) \sgn(l) + 4\pi \ii \sgn(l)\, \delta_{1,\abs{l}} \text{.}
\end{align}
For $I_5$, we find
\begin{align}
I_5 &= \int\!\!\opd\Delta\, \exp{\ii l \Delta}\, \frac{\sin^2\Delta}{\sin^4(\Delta/2)}=  \oint_{\abs{u}=1} f_5(u)\,  \opd{u} \text{,}\qquad f_5(u)=\frac{4 \ii (u+1)^2 u^{l-1}}{(u-1)^2} \text{,}
\end{align}
with $\Res(f_5,1)=16\ii l$ and $\Res(f_5,0)=-16\ii l\, \Theta(-l)+4\ii\, \delta_{0,l}$, which gives
\begin{align}
I_5 &= - 16 \pi l \sgn(l) - 8\pi \delta_{0,l} \text{.}
\end{align}
For $I_6$, we find
\begin{align}
I_6 &= \int\!\!\opd\Delta\, \exp{\ii l \Delta}\, \frac{\sin^2\Delta\cos\Delta}{\sin^4(\Delta/2)}=  \oint_{\abs{u}=1} f_6(u)\,  \opd{u} \text{,}\qquad f_6(u)=\frac{2 \ii (u+1)^2 \left(u^2+1\right) u^{l-2}}{(u-1)^2} \text{,}
\end{align}
with $\Res(f_6,1)=16\ii l$ and $\Res(f_6,0)=-16\ii l\, \Theta(-l)+8\ii\, \delta_{0,l}+2 \ii\, \delta_{1,\abs{l}}$, which gives
\begin{align}
I_6 &= - 16 \pi l \sgn(l) - 16\pi \delta_{0,l} -4\pi \delta_{1,\abs{l}} \text{.}
\end{align}
For $I_7$, we first simplify the integrand with $\sin\Delta=2\sin(\Delta/2)\cos(\Delta/2)$,
\begin{align}
I_7&=\int\!\!\opd\Delta\, \exp{\ii l \Delta}\, \frac{\sin\Delta\cos\Delta\cos(\Delta/2)}{\sin^3(\Delta/2)}=  2 \int\!\!\opd\Delta\, \exp{\ii l \Delta}\, \frac{\cos\Delta\cos^2(\Delta/2)}{\sin^2(\Delta/2)} = \oint_{\abs{u}=1} f_7(u)\,\opd u  \text{,}\\
f_7(u) &= \frac{\ii (u+1)^2 \left(u^2+1\right) u^{l-2}}{(u-1)^2} \text{,}
\end{align}
with $\Res(f_7,1)=8\ii l$ and $\Res(f_7,0)=-8\ii l\, \Theta(-l)+4\ii\,\delta_{0,l}+\ii\, \delta_{1,\abs{l}}$, which gives
\begin{align}
I_7 &= - 8 \pi l \sgn(l) - 8\pi \delta_{0,l} -2\pi \delta_{1,\abs{l}} \text{.}
\end{align}

\subsection{Regularized integral $\tilde I_9$}

In contrast to $I_1$ to $I_8$, the integrand of $I_9$ diverges as $1/\abs{\Delta}$, which is not a pole. The divergence is logarithmic, and its finite part would depend on the scale of the cutoff. We therefore evaluate $I_9$ with an explicit regularization, $\abs{\sin(\Delta/2)}\to\sqrt{\sin^2(\Delta/2)+\delta^2}$,
\begin{align}
\tilde{I}_9(k)= \int \opd{\Delta}\, \frac{\cos(k\Delta) \cos\Delta}{\sqrt{\sin^2(\Delta/2)+\delta^2}} \text{.}
\end{align}
Using $\cos(k\Delta)\cos\Delta=\frac{1}{2} \big(\cos((k+1)\Delta)+\cos((k-1)\Delta)\big)$, this reduces to
\begin{align}
\tilde{I}_9(k) &= \frac{1}{2} \big( C_{k+1}+C_{k-1}\big) \text{,}\qquad
C_k = \int \opd{\Delta}\, \frac{\cos(k\Delta)}{\sqrt{\sin^2(\Delta/2)+\delta^2}} \text{.}
\end{align}
We split $\cos(k\Delta)=1+(\cos(k\Delta)-1)$, which leads to a singular contribution $S$ and a regular part $R_k$, with $\varphi=\Delta/2$,
\begin{align}
C_k &=4  (S + R_{\abs{k}}) \text{,}\qquad
S = \int_0^{\pi/2} \frac{\opd{\varphi}}{\sqrt{\sin^2\varphi+\delta^2}} \text{,}\qquad
R_k = \int_0^{\pi/2} \opd{\varphi}\, \frac{\cos(2k\varphi)-1}{\sqrt{\sin^2\varphi+\delta^2}} \text{,}
\end{align}
where we can safely set $\delta\to 0$ in $R_k$. The singular part is a complete elliptic integral of the first kind~\cite{NIST:DLMF},
\begin{align}
S = \frac{1}{\sqrt{1+\delta^2}}\int_0^{\pi/2} \frac{\opd{\varphi}}{\sqrt{1-c^2 \cos^2\varphi}} = \frac{K(c)}{\sqrt{1+\delta^2}} \text{,}
\end{align}
with $c^2=1/(1+\delta^2)$. For $\delta\to 0$, we are interested in the limit $c\to 1$. With the asymptotic form~\cite{NIST:DLMF} $K(c) \to \log(4/c')$ for $c\to 1^-$, where $c'=\sqrt{1-c^2} \to \delta$, we find
\begin{align}
S = -\log\delta+\log 4 + o(1) \text{.}
\end{align}
For the regular part, we use $(\cos(2k\varphi)-1)/\sin\varphi = -2 \sum_{j=1}^k \sin((2j-1)\varphi)$, which follows from the product formula for sines. This gives
\begin{align}
R_k = - 2 \sum_{j=1}^k \frac{1}{2j-1} \text{,}
\end{align}
so that $R_k \sim -\log k$ for large $k$, which becomes relevant only for high modes at fixed $\Pe^{-1}$. Combining both parts gives \cref{eq:I9reg}.

\end{document}